\documentclass[%
 reprint,
 superscriptaddress,
 nofootinbib,
 amsmath,amssymb,
 aps,
floatfix,
]{revtex4-1}

\usepackage[english]{babel}
\usepackage{color}
\usepackage{graphicx}
\usepackage{epsfig}
\usepackage{amsmath,epsfig}
\usepackage{bm}
\usepackage{amssymb}
\usepackage{subfig}
\usepackage{appendix}
\usepackage{graphicx}
\usepackage{dcolumn}
\usepackage{bm}
\usepackage{txfonts}
\newcommand{\be}{\begin{equation}}
\newcommand{\ee}{\end{equation}}
\newcommand{\ba}{\begin{eqnarray}}
\newcommand{\ea}{\end{eqnarray}}
\newcommand{\nn}{\nonumber\\}
\newcommand{\beq}{\begin{equation}}
\newcommand{\eeq}{\end{equation}}  
\newcommand{\bea}{\begin{eqnarray}}
\newcommand{\eea}{\end{eqnarray}}
\newcommand{\beqa}{\begin{eqnarray}}
\newcommand{\eeqa}{\end{eqnarray}}
\newcommand{\bseq}{\begin{subequations}}
\newcommand{\eseq}{\end{subequations}}

\definecolor{M_Green}        {rgb}{0.03 , 0.6 , 0.35}

\def\thv{{\varpi}}
\def\snn{{\sqrt s_{\rm NN}}}
\def\e{{\rm e}}
\def\di{{\rm d}}

\def\x{{\boldsymbol x}}

\def\0{{\boldsymbol 0}}

\def\cal{\mathcal}
\def\w{{\tilde\omega}}
\def\hs{\hat{s}}
\def\hx{\hat{x}}

\def\hT{\hat{T}}
\def\hu{\hat{u}}

\begin{document}

\title{A study of vorticity formation in high energy nuclear collisions} 

\author{F.~Becattini}
\affiliation{Dipartimento di Fisica e Astronomia, Universit\`a di Firenze, 
Via G. Sansone 1, I-50019 Sesto F.no (Firenze), Italy}
\affiliation{INFN - Sezione di Firenze, Via G. Sansone 1, I-50019 Sesto F.no (Firenze), Italy}

\author{G.~Inghirami}
\affiliation{Frankfurt Institute for Advanced Studies (FIAS), 
Johann Wolfgang Goethe University, Frankfurt am Main, Germany}
\affiliation{Dipartimento di Fisica e Astronomia, Universit\`a di Firenze, 
Via G. Sansone 1, I-50019 Sesto F.no (Firenze), Italy}

\author{V.~Rolando}
\affiliation{Dipartimento di Fisica e Scienze della Terra, Universit\`a di Ferrara, 
  Via Saragat 1, I-44100 Ferrara, Italy}
\affiliation{INFN - Sezione di Ferrara, Via Saragat 1, I-44100 Ferrara, Italy}

\author{A.~Beraudo}
\affiliation{INFN - Sezione di Torino, Via P. Giuria 1, I-10125 Torino, Italy}

\author{L.~\surname{Del~Zanna}}
\affiliation{Dipartimento di Fisica e Astronomia, Universit\`a di Firenze, 
Via G. Sansone 1, I-50019 Sesto F.no (Firenze), Italy}
\affiliation{INFN - Sezione di Firenze, Via G. Sansone 1, I-50019 Sesto F.no (Firenze), Italy}
\affiliation{INAF - Osservatorio Astrofisico di Arcetri, L.go E. Fermi 5, I-50125 Firenze, Italy}

\author{A.~\surname{De Pace}}
\affiliation{INFN - Sezione di Torino, Via P. Giuria 1, I-10125 Torino, Italy}

\author{M.~\surname{Nardi}}
\affiliation{INFN - Sezione di Torino, Via P. Giuria 1, I-10125 Torino, Italy}

\author{G.~Pagliara}
\affiliation{Dipartimento di Fisica e Scienze della Terra, Universit\`a di Ferrara, 
  Via Saragat 1, I-44100 Ferrara, Italy}
\affiliation{INFN - Sezione di Ferrara, Via Saragat 1, I-44100 Ferrara, Italy}

\author{V.~Chandra}
\affiliation{Indian  Institute of Technology Gandhinagar, Ahmedabad-382424, Gujrat, 
India}

\date{\today}

\begin{abstract}
We present a quantitative study of vorticity formation in peripheral ultrarelativistic 
heavy ion collisions at $\snn = 200$ GeV by using the ECHO-QGP numerical code, 
implementing relativistic dissipative hydrodynamics in the causal Israel-Stewart 
framework in 3+1 dimensions with an initial Bjorken flow profile. We consider and
discuss different definitions of vorticity which are relevant in relativistic 
hydrodynamics. After demonstrating the excellent capabilities of our code, which
proves to be able to reproduce Gubser flow up to 8 fm/$c$, we show that, with the 
initial conditions needed to reproduce the measured directed flow in peripheral 
collisions corresponding to an average impact parameter $b=11.6$ fm 
and with the Bjorken flow profile for a viscous Quark Gluon Plasma with $\eta/s=0.1$ 
fixed, a vorticity of the order of some $10^{-2} \; c$/fm can develop at freezeout. 
The ensuing polarization of $\Lambda$ baryons does not exceed 1.4\% at midrapidity. 
We show that the amount of developed directed flow is sensitive to both the initial 
angular momentum of the plasma and its viscosity.
\end{abstract}


\maketitle

\section{Introduction}

The hydrodynamical model has by now become a paradigm for the study of the QCD plasma
formed in nuclear collisions at ultrarelativistic energies. There has been a considerable 
advance in hydrodynamics modeling and calculations of these collisions over the 
last decade. Numerical simulations in 2+1D \cite{vishnu} and in 3+1 D \cite{music,
echoqgp,karpe,molnar,japan,bozekcode} including viscous corrections are becoming the new 
standard in this field and existing codes are also able to handle initial state fluctuations. 

An interesting issue is the possible formation of vorticity in peripheral collisions 
\cite{beca0,csernai1,csernai2}. Indeed, the presence of vorticity may provide information about 
the (mean) initial state of the hydrodynamical evolution which cannot be achieved 
otherwise, and it is related to the onset of peculiar physics in the plasma at high 
temperature, such as the chiral vortical effect \cite{chiral}. Furthermore, it has 
been shown that vorticity gives rise to polarization of particles in the final state, 
so that e.g. $\Lambda$ baryon polarization - if measurable - can be used to detect 
it \cite{beca1,beca2}. 
Finally, as we will show, numerical calculation of vorticity can be used to make 
stringent tests of numerical codes, as the T-vorticity (see sect.~\ref{defin} for the 
definition) is expected to vanish throughout under special initial conditions in the 
ideal case. 

Lately, vorticity has been the subject of investigations in refs.~\cite{csernai1,
csernai2} with peculiar initial conditions in cartesian coordinates, ideal fluid 
approximation and isochronous freezeout. Instead, in this work, we calculate different 
kinds of vorticity with our 3+1D ECHO-QGP \footnote{The code is publicly available 
at the web site http://theory.fi.infn.it/echoqgp} code \cite{echoqgp}, including dissipative 
relativistic hydrodynamics in the Israel-Stewart formulation with Bjorken initial 
conditions for the flow (i.e. with $u^x=u^y=u^\eta=0$), henceforth denoted as BIC.
It should be pointed out from the very beginning that the purpose of this work is 
to make a general assessment of vorticity at top RHIC energy and not to provide a 
precision fit to all the available data. Therefore, our calculations do not take 
into account effects such as viscous corrections to particle distribution at the 
freezeout and initial state fluctuations, that is we use smooth initial conditions
obtained averaging over many events.

\subsection{Notations}
In this paper we use the natural units, with $\hbar=c=K=1$.\\ 
The Minkowskian metric tensor is ${\rm diag}(1,-1,-1,-1)$; for the Levi-Civita
symbol we use the convention $\epsilon^{0123}=1$.\\ 
We will use the relativistic notation with repeated indices assumed to 
be summed over, however contractions of indices will be sometimes denoted with 
dots, e.g. $ u \cdot T \cdot u \equiv u_\mu T^{\mu\nu} u_\nu$.
The covariant derivative is denoted as $d_\mu$ (hence $d_\lambda g_{\mu\nu}=0$), 
the exterior derivative by ${\bf d}$, whereas $\partial_\mu$ is the ordinary 
derivative. 

\section{Vorticities in relativistic hydrodynamics}
\label{defin}

Unlike in classical hydrodynamics, where vorticity is the curl of the velocity field 
${\bf v}$, several vorticities can be defined in relativistic hydrodynamics which 
can be useful in different applications (see also the review~\cite{gourg}). 

\subsection{The kinematical vorticity}

This is defined as:
\be\label{kinvort}
  \omega_{\mu\nu} = \frac{1}{2} (d_\nu u_\mu - d_\mu u_\nu) = 
  \frac{1}{2} (\partial_\nu u_\mu - \partial_\mu u_\nu)
\ee
where $u$ is the four-velocity field. This tensor includes both the acceleration 
$A$ and the relativistic extension of the angular velocity pseudo-vector $\omega_\mu$ 
in the usual decomposition of an antisymmetric tensor field into a polar and 
pseudo-vector fields:
\bea\label{decomp}
   \omega_{\mu\nu} &=& \epsilon_{\mu\nu\rho\sigma} \omega^\rho u^\sigma
   + \frac{1}{2} (A_\mu u_\nu - A_\nu u_\mu)  \nn 
   A_\mu &=& 2 \omega_{\mu\nu} u^\nu = u^\nu d_\nu u_\mu \equiv D u_\mu  \nn 
   \omega_\mu &=& -\frac{1}{2} \epsilon_{\mu\rho\sigma\tau} \,\omega^{\rho\sigma} 
   u^\tau
\eea
where $\epsilon_{\mu\nu\rho\sigma}$ is the Levi-Civita symbol.
Using of the transverse (to $u$) projector:
$$
 \Delta^{\mu\nu}\equiv g^{\mu\nu} - u^{\mu}u^{\nu},
$$
and the usual definition of the orthogonal derivative
$$
  \nabla_\mu \equiv  \Delta^\alpha_\mu d_\alpha = d_\mu - u_\mu D,
$$
where $D=u^\alpha d_\alpha$, it is convenient to define also a transverse kinematical 
vorticity as:
\be
  \omega^{\Delta}_{\mu\nu} = \Delta_{\mu\rho} \Delta_{\nu\sigma} 
  \omega^{\rho\sigma} = \frac{1}{2}(\nabla_\nu u_\mu - \nabla_\mu u_\nu)
\ee
Using the above definition in the decomposition (\ref{decomp}) it can be shown that:
\be
 \omega^{\Delta}_{\mu\nu} = \epsilon_{\mu\nu\rho\sigma} \omega^\rho u^\sigma
\ee
that is $\omega^\Delta$ is the tensor formed with the angular velocity vector only.
As we will show in the next subsection, only $\omega^\Delta$ shares the ``conservation"
property of the classical vorticity for an ideal barotropic fluid.

\subsection{The T-vorticity}

This is defined as:
\be\label{tvort}
  \Omega_{\mu\nu} = \frac{1}{2} \left[ \partial_\nu (T u_\mu) - \partial_\mu 
  (T u_\nu) \right]
\ee
and it is particularly useful for a relativistic uncharged fluid, such as the QCD
plasma formed in nuclear collisions at very high energy. This is because from the 
basic thermodynamic relations when the temperature is the only independent 
thermodynamic variable, the ideal relativistic equation of motion $(\varepsilon + p) 
A_\mu = \nabla_\mu p$ can be recast in the simple form (see e.g. \cite{stephanov}):
\be\label{carter}
  u^\mu \Omega_{\mu\nu} = \frac{1}{2} (T A_\nu - \nabla_\nu T)  = 0 
\ee     

The above (\ref{carter}) is also known as Carter-Lichnerowicz equation \cite{gourg} for 
an ideal uncharged fluid and it entails conservation properties which do not hold for the 
kinematical vorticity. This can be better seen in the the language of differential forms, 
rewriting the definition of the T-vorticity as the exterior derivative of a the vector 
field (1-form) $Tu$, that is $\Omega = \mathbf{d} (Tu)$. Indeed, the eq.~(\ref{carter})
implies - through the Cartan identity - that the Lie derivative of $\Omega$ along 
the vector field $u$ vanishes, that is
\be\label{cartan}
  {\cal L}_u \,\Omega = u \cdot {\bf d} \Omega + {\bf d} (u \cdot \Omega) = 0
\ee
because $\Omega$ is itself the external derivative of the vector field $T u$ and
${\bf d}{\bf d} = 0$. The eq.~(\ref{cartan}) states that the T-vorticity is conserved
along the flow and, thus, if it vanishes at an initial time it will remain so at
all times. This can be made more apparent by expanding the Lie derivative definition
in components:
\be\label{lie}
 ({\cal L}_u \,\Omega)^{\mu\nu} = D \Omega^{\mu\nu} - \partial_\sigma u^\mu
 \Omega^{\sigma\nu} - \partial_\sigma u^\nu \Omega^{\sigma\mu} = 0
\ee 
The above equation is in fact a differential equation for $\Omega$ precisely showing 
that if $\Omega=0$ at the initial time then $\Omega\equiv 0$. Thereby, the T-vorticity
has the same property as the classical vorticity for an ideal barotropic fluid, such 
as the Kelvin circulation theorem, so the integral of $\Omega$ over a surface enclosed 
by a circuit comoving with the fluid will be a constant.

One can write the relation between T-vorticity and kinematical vorticity by expanding
the definition (\ref{tvort}):
$$
  \Omega_{\mu\nu} = \frac{1}{2}\left[ (\partial_\nu T) \, u_\mu - (\partial_\mu T) \,
  u_\nu \right] + T \omega_{\mu\nu}
$$
implying that the double-transverse projection of $\Omega$:
$$
  \Delta_{\mu\rho}\Delta_{\nu\sigma} \Omega^{\rho\sigma} \equiv \Omega^\Delta_{\mu\nu}
  = T \omega_{\mu\nu}^\Delta 
$$
Hence, the tensor $\omega^\Delta$ shares the same conservation properties of $\Omega^\Delta$,
namely it vanishes at all times if it is vanishing at the initial time. Conversely, 
the mixed projection of the kinematical vorticity:
$$
  u^\rho \omega_{\rho\sigma} \Delta^{\sigma\nu} = \frac{1}{2} A_{\sigma}
$$
does not. It then follows that for an ideal uncharged fluid with $\omega^\Delta=0$
at the initial time, the kinematical vorticity is simply:
\be\label{omeglong}
  \omega_{\mu\nu} = \frac{1}{2} ( A_\mu u_\nu - A_\nu u_\mu)
\ee
%

\subsection{The thermal vorticity}

This is defined as \cite{beca2}:
\be\label{thvort}
  \thv_{\mu\nu} = \frac{1}{2} (\partial_\nu \beta_\mu - \partial_\mu \beta_\nu)   
\ee
where $\beta$ is the temperature four-vector. This vector is defined as $(1/T) u$
once a four-velocity $u$, that is a hydrodynamical frame, is introduced, but it
can also be taken as a primordial quantity to define a velocity through $u \equiv
\beta/\sqrt{\beta^2}$ \cite{becaframe}. The thermal vorticity features two 
important properties: it is adimensional in natural units (in cartesian coordinates)
and it is the actual constant vorticity at the global equilibrium with rotation 
\cite{becacov} for a relativistic system, where $\beta$ is a Killing vector field 
whose expression in Minkowski spacetime is $\beta_\mu = b_\mu + \thv_{\mu\nu} x^\nu$ 
being $b$ and $\thv$ constant. In this case the magnitude of thermal vorticity is 
- with the natural constants restored - simply $\hbar \omega/k_BT$ where $\omega$ is 
a constant angular velocity. In general, (replacing $\omega$ with the classical 
vorticity defined as the curl of a proper velocity field) it can be readily realized 
that the adimensional thermal vorticity is a tiny number for most hydrodynamical 
systems, though it can be significant for the plasma formed in relativistic nuclear 
collisions. 

Furthermore, the thermal vorticity is responsible for the local polarization of particles 
in the fluid according to the formula \cite{beca1}:
\be
  \Pi_\mu(x,p) = -\frac{1}{8} \epsilon_{\mu\rho\sigma\tau} (1-n_F) \,\,
  \thv^{\rho\sigma} \frac{p^\tau}{m}
\ee
which applies to spin 1/2 fermions, $n_F$ being the Fermi-Dirac-Juttner distribution
function. 
\be\label{juttner}
   n_F = \frac{1}{\e^{\beta(x)\cdot p - \mu/T}+1}
\ee

Similarly to the previous subsection, one can readily obtain the relation between
T-vorticity and thermal vorticity:
\be\label{relaz}
  \thv_{\mu\nu} = \frac{1}{2T^2} \left[ (\partial_\mu T) \, u_\nu - 
  (\partial_\nu T) \, u_\mu \right] + \frac{1}{T^2} \Omega_{\mu\nu}
\ee
Again, the double transverse projection of $\thv$ is proportional to the one of
$\Omega$:
$$
  \Delta_{\mu\rho}\Delta_{\nu\sigma} \thv^{\rho\sigma} \equiv \thv^\Delta_{\mu\nu}
  = \frac{1}{T^2} \Omega_{\mu\nu}^\Delta = \frac{1}{T} \omega^\Delta
$$
whereas the mixed projection turns out to be, using eq.~(\ref{relaz})
$$
  u^\rho \thv_{\rho\sigma} \Delta^{\sigma\nu} = \frac{1}{2T^2} \nabla^\nu T
  + \frac{A^\nu}{2T}
$$
Again, for an ideal uncharged fluid with $\omega^\Delta=0$ at the initial time, 
by using the equations of motion (\ref{carter}), one has the above projection is
just $A^\nu/T$ and that the thermal vorticity is simply:
\be\label{thlong}
  \thv_{\mu\nu} = \frac{1}{T} (A_\mu u_\nu - A_\nu u_\mu)
\ee
A common feature of the kinematical and thermal vorticity is that their purely spatial 
components can be non-vanishing if the acceleration and velocity field are non-parallel, 
even though velocity is vanishing at the beginning.

\section{High energy nuclear collisions}
\label{nuclear}

In nuclear collisions at very large energy, the QCD plasma is an almost uncharged
fluid. Therefore, according to previous section's arguments, in the ideal fluid 
approximation, if the transversely projected vorticity tensor $\omega^\Delta$ initially 
vanishes, so will the transverse projection $\Omega^\Delta$ and $\thv^\Delta$ and
the kinematical and thermal vorticities will be given by the formulae (\ref{omeglong}) 
and (\ref{thlong}) respectively. Indeed, the T-vorticity $\Omega$ will vanish throughout
because also its longitudinal projection vanishes according to eq.~(\ref{carter}). 
This is precisely what happens for the usually assumed BIC for the flow at $\tau_0$, 
that is $u^x=u^y=u^\eta=0$, where one has $\omega^\Delta=0$ at the beginning as it 
can be readily realized from the definition (\ref{kinvort}). On the other hand, for 
a viscous uncharged fluid, transverse vorticities can develop even if they are zero
at the beginning. 

It should be noted though, that even if the space-space components ($x,y,\eta$ indices) 
of the kinematical vorticity tensor vanish at the initial Bjorken time $\tau_0$, they 
can develop at later times even for an ideal fluid if the spatial parts of the acceleration and 
velocity fields are not parallel, according to eq.~(\ref{omeglong}). The equation makes 
it clear that the onset of spatial components of the vorticity is indeed a relativistic
effect as, with the proper dimensions, it goes like (${\bf a} \times {\bf v})/c^2$. 

\begin{figure}
\includegraphics[width=0.5\textwidth]{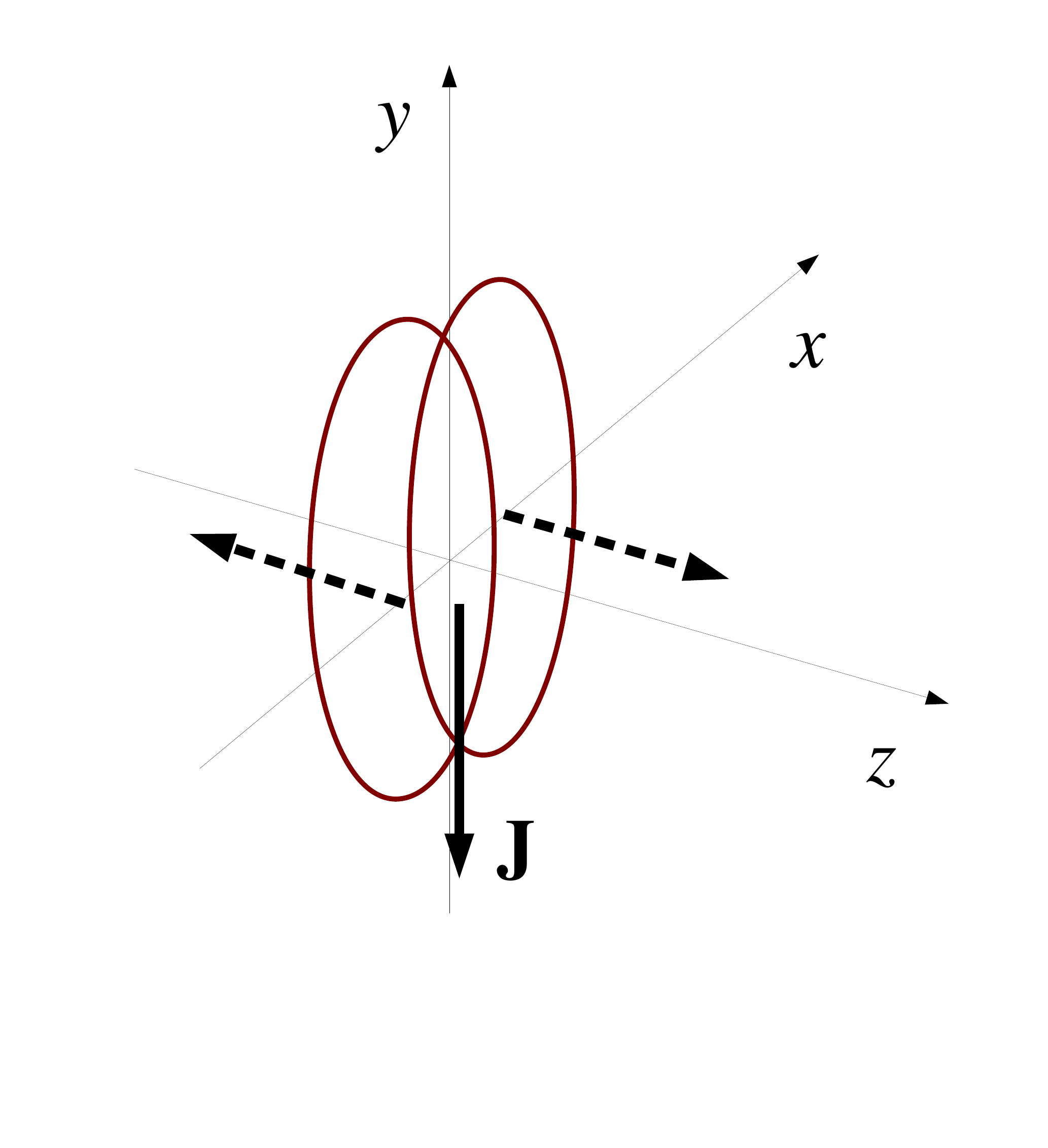}
\caption{Colliding nuclei and conventional cartesian reference frame. Also shown 
the initial angular momentum vector.}
\label{frame}
\end{figure}

In the full longitudinally boost invariant Bjorken picture, that is $u^\eta = 0$ 
throughout the fluid evolution, in the ideal case, as $\omega^\Delta=0$, the only 
allowed components of the kinematical vorticity are $\omega^{\tau x}, \omega^{\tau y}$ 
and $\omega^{xy}$ from the first eq.~(\ref{decomp}). The $\omega^{xy}$ component, 
at $\eta=0$, because of the reflection symmetry (see fig.~\ref{frame}) in both the 
$x$ and $y$ axes, can be different from zero but it ought to change sign by moving 
clockwise (or counterclockwise) to the neighbouring quadrant of the $xy$ plane; for 
central collisions it simply vanishes.  

However, in the viscous case, more components of the vorticities can be non-vanishing. 
Furthermore, in more realistic 3+1 D hydrodynamical calculations, a non-vanishing 
$u^\eta$ can develop because of the asymmetries of the initial energy density in the 
$x-\eta$ and $y-\eta$ planes at finite impact parameter. The asymmetry is essential 
to reproduce the observed directed flow coefficient $v_1(y)$ in a 3+1D ideal hydrodynamic 
calculation with BIC, as shown by Bozek \cite{bozek}, and gives the plasma a total 
angular momentum, as it will be discussed later on. 

In this work, we calculate the vorticities, and especially the thermal vorticity 
$\thv$ by using basically the same parametrization of the initial conditions in 
ref.~\cite{bozek}. Those initial conditions are a modification of the usual BIC 
to take into account that the plasma, in peripheral collisions, has a relatively 
large angular momentum (see Appendix A). They are a minimal modifications of the 
BIC in that the initial flow velocity Bjorken components are still zero, but the 
energy density longitudinal profile is changed and no longer symmetric by the reflection 
$\eta \to -\eta$. They are summarized hereinafter. Given the usual thickness function 
expression: 
\be
T(x,y) =\int_{-\infty}^{\infty} \di z \, n(x,y,z) = \int_{-\infty}^{\infty} \di z \, 
\frac{n_0}{1+\e^{(\sqrt {x^2+y^2+z^2}- R)/\delta}} \ee
where $n_0=0.1693\,\textrm{fm}^{-3}$, $\delta=0.535\,\textrm{fm}$ and $R=6.38\,\textrm{fm}$ 
are the nuclear density, the width and the radius of the nuclear Fermi distribution 
respectively, the following functions are defined:
\bea
 T_1 &=& T_+\,\left (1-\left ( 1-\dfrac{\sigma T_-}{A}\right )^A \right ) \\
 T_2 &=& T_-\,\left (1-\left ( 1-\dfrac{\sigma T_+}{A}\right )^A \right )
\eea
where $\sigma$ is the inelastic NN cross section, $A$ the mass number of the colliding 
nuclei, and:
\be\label{thick}
T_+({\bf x}_T) = T({\bf x}_T+{\bf b}/2)\qquad T_-({\bf x}_T) = T({\bf x}_T-{\bf b}/2)
\ee
where ${\bf x}_T=(x,y)$ is the vector of the transverse plane coordinates and 
${\bf b}$ is the impact parameter vector, connecting the centers of the two nuclei. 
In our conventional cartesian reference frame, the ${\bf b}$ vector is oriented along 
the positive $x$ axis and the two nuclei have initial momentum along the $z$ axis 
(whence the reaction plane is the $xz$ plane) and their momenta are directed so as to 
make the initial total angular momentum oriented along the negative $y$ axis (see
fig.~\ref{frame}). The wounded nucleons weight function $W_N$ is then defined:
\be\label{wound}
 W_N(x,y,\eta)=2\,\left( T_1(x,y)f_{-}(\eta)+T_2(x,y)f_{+}(\eta)\right)
\ee
where:
$$
f_{-}(\eta)=
\begin{cases}
1 & \eta < -\eta_m\\
\dfrac{-\eta+\eta_m}{2\eta_m} & -\eta_m \le \eta \le \eta_m\\
0 & \eta > \eta_m
\end{cases}
$$
and
$$
f_{+}(\eta)=
\begin{cases}
0 & \eta < -\eta_m\\
\dfrac{\eta+\eta_m}{2\eta_m} & -\eta_m \le \eta \le \eta_m\\
1 & \eta > \eta_m
\end{cases}
$$
Finally, the initial proper energy density distribution is assumed to be:
\be\label{inidens}
 \varepsilon(x,y,\eta)=\varepsilon_0 \,W(x,y,\eta)\,H(\eta),
\ee
where the total weight function $W(x,y,\eta)$ is defined as:
\be\label{weight}
W(x,y,\eta)=\dfrac{(1-\alpha)\,W_{N}(x,y,\eta)+\alpha\,n_{BC}(x,y)}{(1-\alpha)
\,W_{N}(0,0,0)+\alpha\,n_{BC}(0,0)\Big|_{{\bf b}=0}}.
\ee
and: 
\be\label{profile}
  H(\eta)=\exp\left(-\dfrac{\tilde\eta^2}{2\sigma_{\eta}^2}\theta(\tilde \eta)\right)
  \qquad  \tilde\eta=|\eta|-\eta_{flat}/2
\ee
In the eq.~(\ref{weight}) $n_{BC}(x,y)$ is the mean number of binary collisions:
\be
n_{BC}(x,y)={\sigma}_{in} T_+(x,y)\,T_-(x,y)
\ee
and $\alpha$ is the {\em collision hardness} parameter, which can vary between 
0 and 1. 

This parametrization, and especially the chosen forms of the functions $f_\pm$, are 
certainly not unique as a given angular momentum can be imparted to the plasma in 
infinitely many ways. Nevertheless, as has been mentioned, it proved to reproduce 
correctly the directed flow in a 3+1D hydrodynamical calculation of peripheral Au-Au
collisions at high energy \cite{bozek}, thus we took it as a good starting point. A 
variation of this initial condition will be briefly discussed in sect.~\ref{conclu}.
Besides, the parametrization (\ref{inidens}) essentially respects the causality constraint 
that the plasma cannot extend beyond $\eta = y_{\rm beam}$. Indeed, at $\snn =200$ GeV 
$y_{\rm beam}\simeq 5.36$ while the 3 $\sigma$ point in the gaussian profile in 
eq.~(\ref{profile}) lies at $\eta = \eta_{\rm flat}/2 + 3 \sigma_\eta \simeq 4.4$. 

The free parameters have been chosen following ref.~\cite{heinzic}, where they 
were adjusted to reproduce the data in Au-Au collisions at $\snn =200$ GeV.
They are reported in table~\ref{partable}.
\begin{table}[!ht]\begin{center}
\begin{tabular}{|c|c|}
\hline
Parameter &Value \\
\hline
$\sqrt{s_{NN}}$&$200\,\textrm{GeV}$\\
$\alpha$&$0.$\\
$\epsilon_0$&$30\,\textrm{GeV}/\textrm{fm}^\textrm{3}$\\
$\sigma_{in}$&$40\,\textrm{mb}$\\
$\tau_0$&$0.6\,\textrm{fm}/{c}$\\
$\eta_{flat}$&$1$\\
$\sigma_{\eta}$&$1.3$\\
$T_{fo}$&$130\, \textrm{MeV}$\\
$b$& $11.57\textrm{fm}$\\
\hline
$\eta_m$ ideal & 3.36 \\
$\eta_m$ viscous & 2.0 \\
$\eta/s$ & 0.1 \\
\hline
\end{tabular}\end{center}
\caption{Parameters defining the initial configuration of the fluid in 
Bjorken coordinates. The last two parameter values have been fixed for the
last physical run.}
\label{partable}
\end{table}

We have run the ECHO-QGP code in both the ideal and viscous modes with the parameters
reported in table~\ref{partable} and the equation of state reported in ref.~\cite{laine}. 
The impact parameter value $b=11.57$ was chosen as, in the optical Glauber model, it 
corresponds to the mean value of the 40-80\% centrality class ($9.49 < b < 13.42$ fm 
\cite{steinberg}) used by the STAR experiment for the directed flow measurement in ref.
~\cite{star08}. 
The initial flow velocities $u^x,u^y,u^\eta$ were set to zero, according to BIC. The 
freezeout hypersurface - isothermal at $T_{fo}=130$ MeV - is determined with the 
methods described in refs.~\cite{echoqgp,rolando}.

\section{Qualification of the ECHO-QGP code}

To show that our code is well suited to model the evolution of the matter produced 
in heavy-ion collisions and hence to carry out our study on the development of vorticity 
in such an environment, we have performed two calculations, referring to an ideal and 
viscous scenario respectively, providing a very stringent numerical test. 

Before describing these tests, it should be pointed out that the vorticities components 
are to be calculated in Bjorken coordinates, whose metric tensor is $g_{\mu\nu} = 
\mathrm{diag}(1,-1,-1,-\tau^2)$, hence they do not all have the same dimension nor 
they are adimensional as it is desirable (except the thermal vorticity, as it has 
been emphasized in Sect.~\ref{defin}). For a proper comparison it is better to use 
the orthonormal basis, which involves a factor $\tau$ when the $\eta$ components 
are considered. Moreover, the cumulative contribution of all components is well 
described by the invariant modulus, which, for a generic antisymmetric tensor 
$A_{\mu\nu}$ is:
\be
 A^2 = A_{\mu\nu}A^{\mu\nu} = 2[A_{xy}^2 - A_{\tau x}^2 - A_{\tau y}^2 + 
( A_{\eta x}^2 + A_{\eta y}^2 - A_{\eta\tau}^2)/\tau^2].
\ee
Furthermore, we have always rescaled the T-vorticity by $1/T^2$ in order to have 
an adimensional number. Since the T-vorticity has always been determined at the 
isothermal freezeout, in order to get its actual magnitude, one just needs to 
multiply it by $T_{fo}^2$.

\subsection{T-vorticity for an ideal fluid}

\begin{figure}[!ht]
\includegraphics[width=0.5\textwidth]{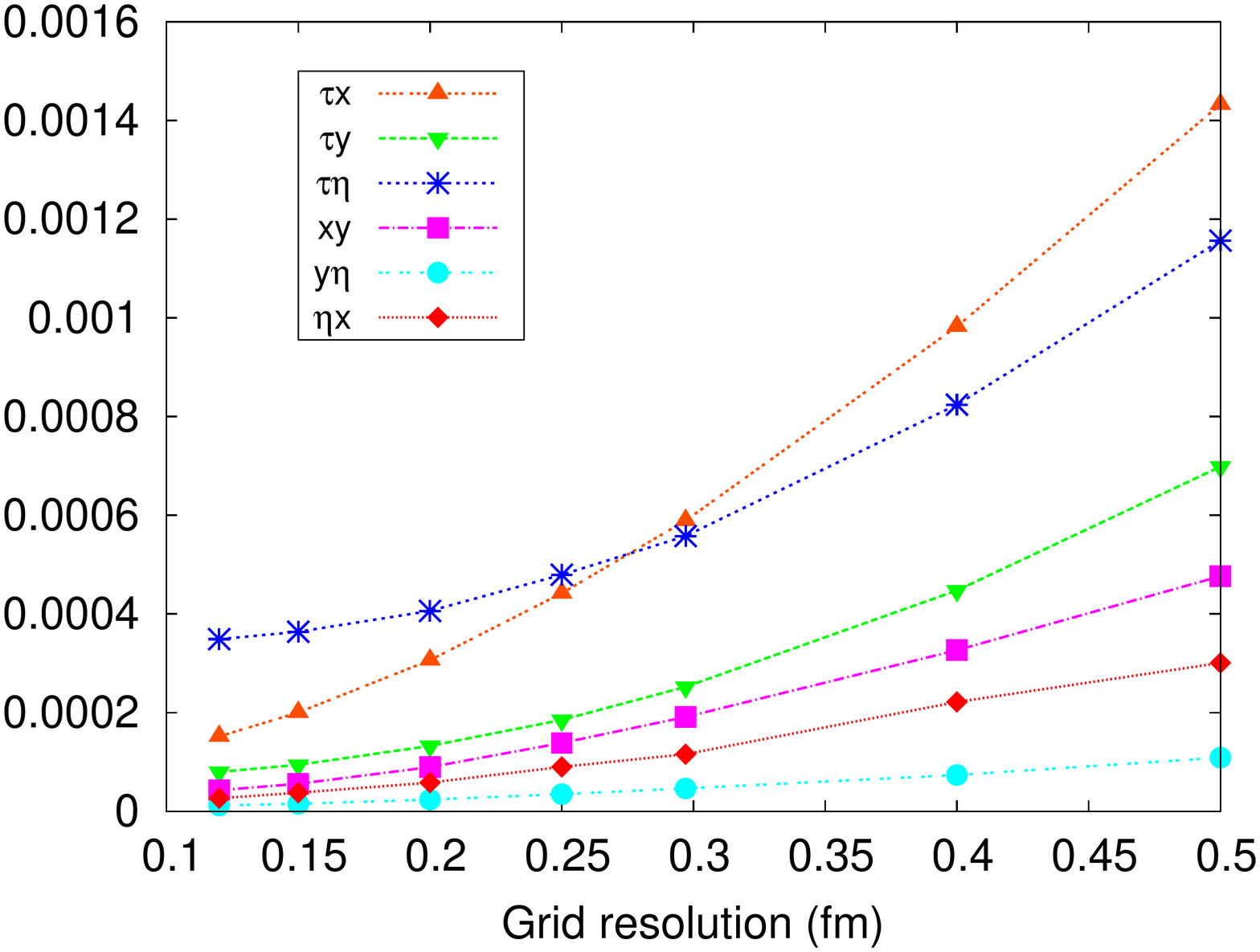}
\caption{(color online) Mean of the absolute value of T-vorticity components, 
divided by $T^2$, at the freeze-out as a function of the grid resolution.}
\label{resol}
\end{figure}

Since the fluid is assumed to be uncharged and the initial T-vorticity $\Omega$ 
is vanishing with the BIC, it should be vanishing throughout, according to the 
discussion in sect.~\ref{defin}). However, the discretization of the hydrodynamical 
equations entails a numerical error, thus the smallness of $\Omega$ in an ideal run 
is a gauge of the quality of the computing method. In fig.~\ref{resol} we show the 
mean of the absolute values of the six independent Bjorken components at the freeze-out 
hypersurface, of the T-vorticity divided by $T^2$ to make it adimensional, as 
a function of the grid resolution (the boundaries in $x,y,\eta$ being fixed) 
\footnote{It should be pointed out that, throughout this work, by mean values of 
the vorticities we mean simple averages of the (possibly rescaled by $1/\tau$) Bjorken 
components over the freezeout hypersurface without geometrical cell weighting. 
Therefore, the plotted mean values have no physical meaning and they should be 
taken as descriptive numbers which are related to the global features of vorticity
components at the freeze-out.}
As it is expected, the normalized T-vorticity decreases as the resolution improves.   

Because of the relation (\ref{relaz}), the residual value at our best spatial resolution 
of 0.15 fm can be taken as a numerical error for later calculations of the thermal 
vorticity.

\begin{figure}
\includegraphics[width=0.5\textwidth]{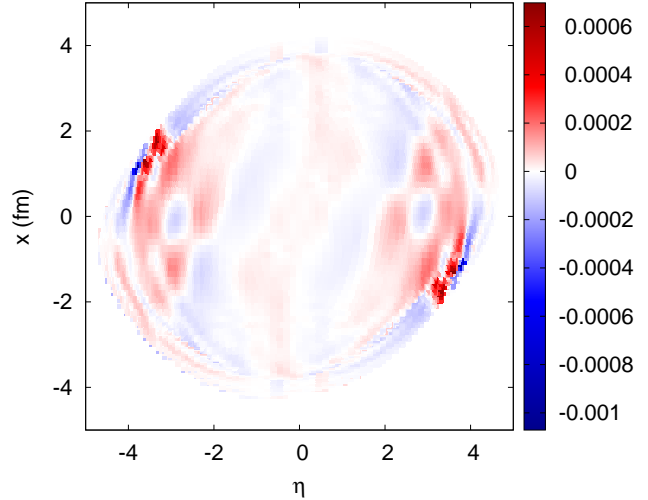}
\caption{(color online) Contour plot of $\Omega_{x\eta}/\tau T^2$ at the freeze-out 
hypersurface at $y=0$.}
\label{tvmap}
\end{figure}

\subsection{Gubser flow}

A very useful test for the validation of a numerical code of relativistic dissipative 
hydrodynamics is the explicit solution of Israel-Stewart theory of a Bjorken flow 
with an azimuthally symmetric radial expansion~\cite{gub1,gub2,den1,den2}, the so-called
Gubser flow. Indeed, this solution provides a highly non-trivial theoretical benchmark.
\begin{figure*}[!ht]
 \includegraphics[width=0.5\textwidth]{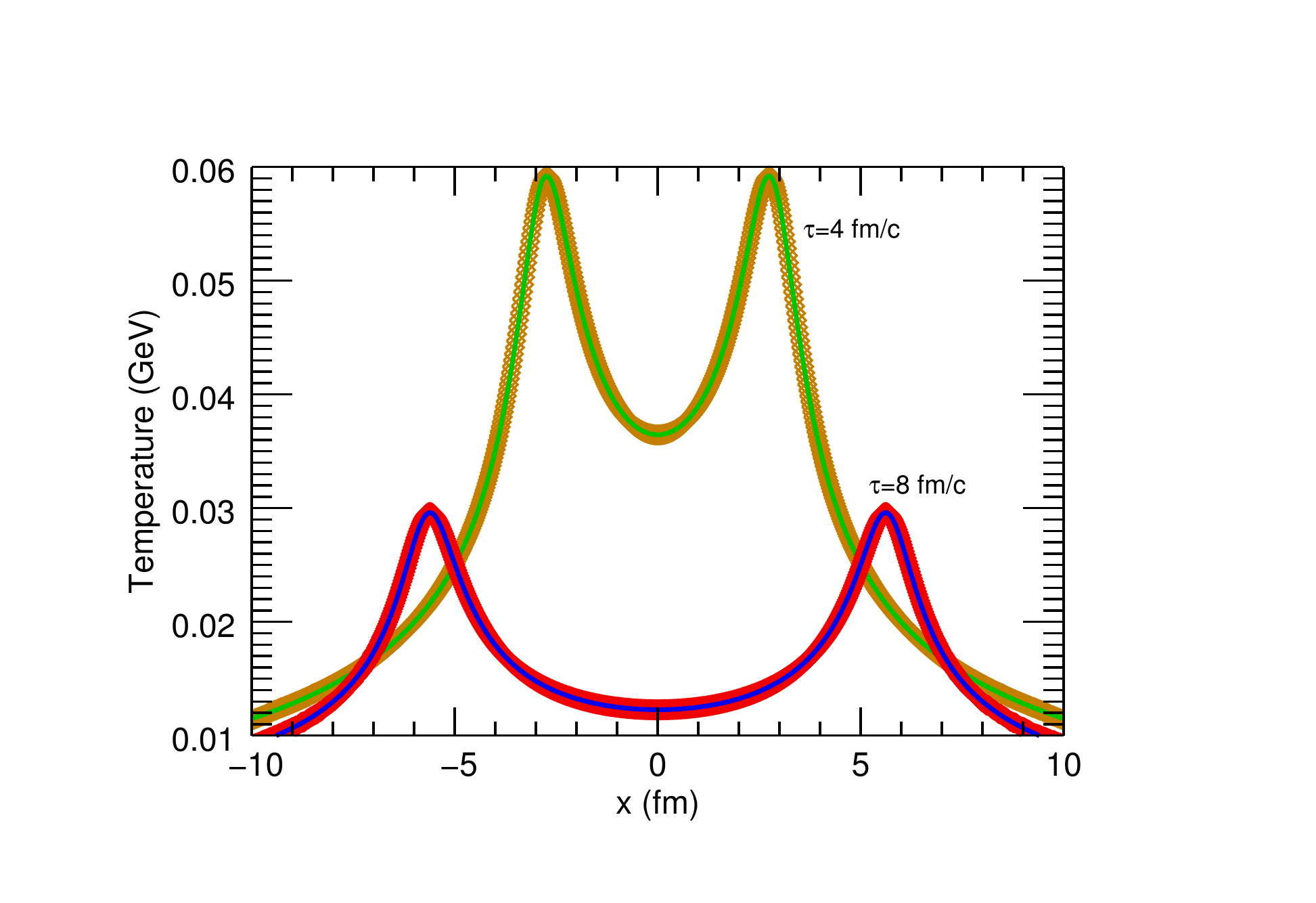}
 \hspace{-15mm}
 \includegraphics[width=0.5\textwidth]{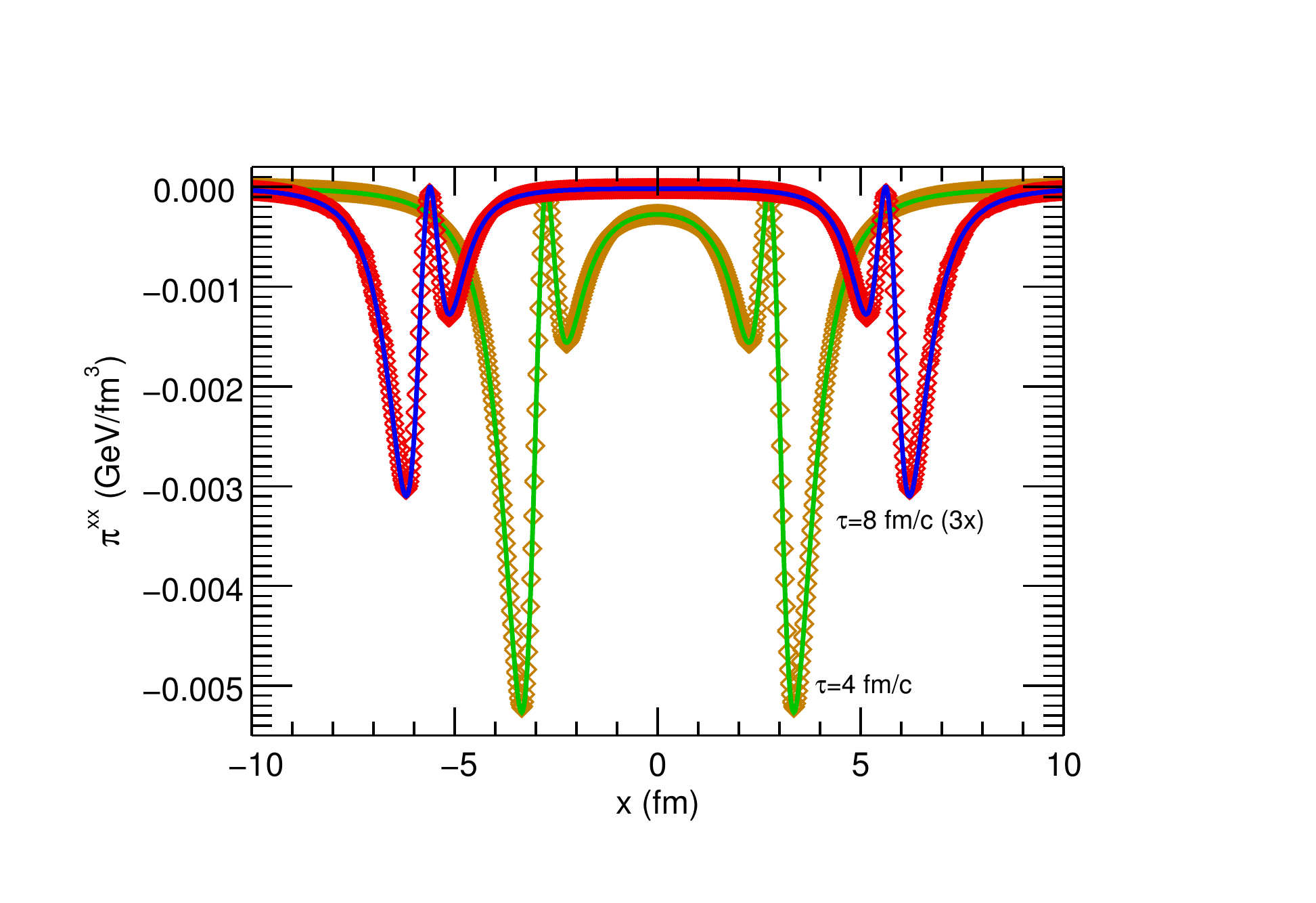}
      
 \includegraphics[width=0.5\textwidth]{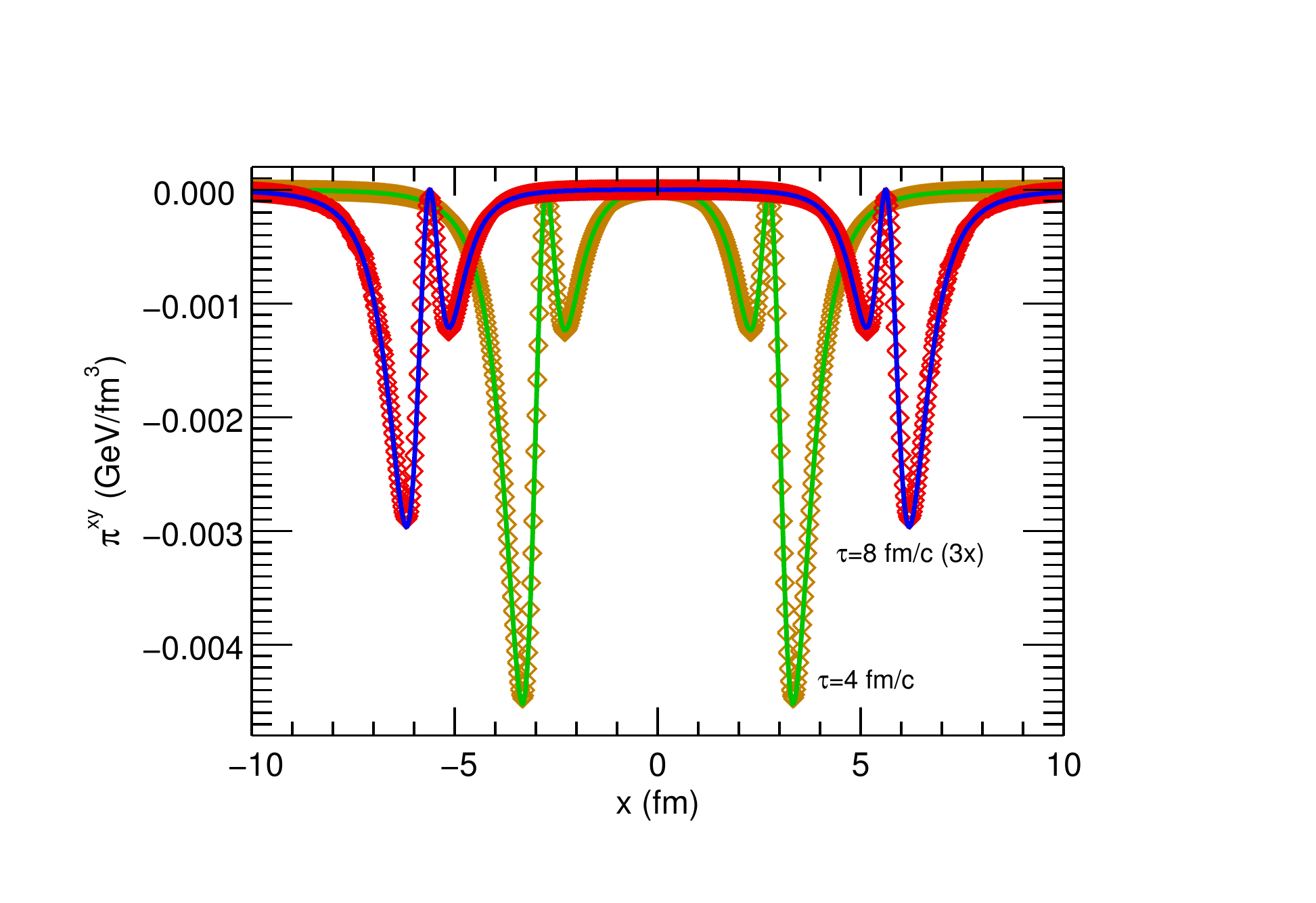}
\hspace{-15mm}
 \includegraphics[width=0.5\textwidth]{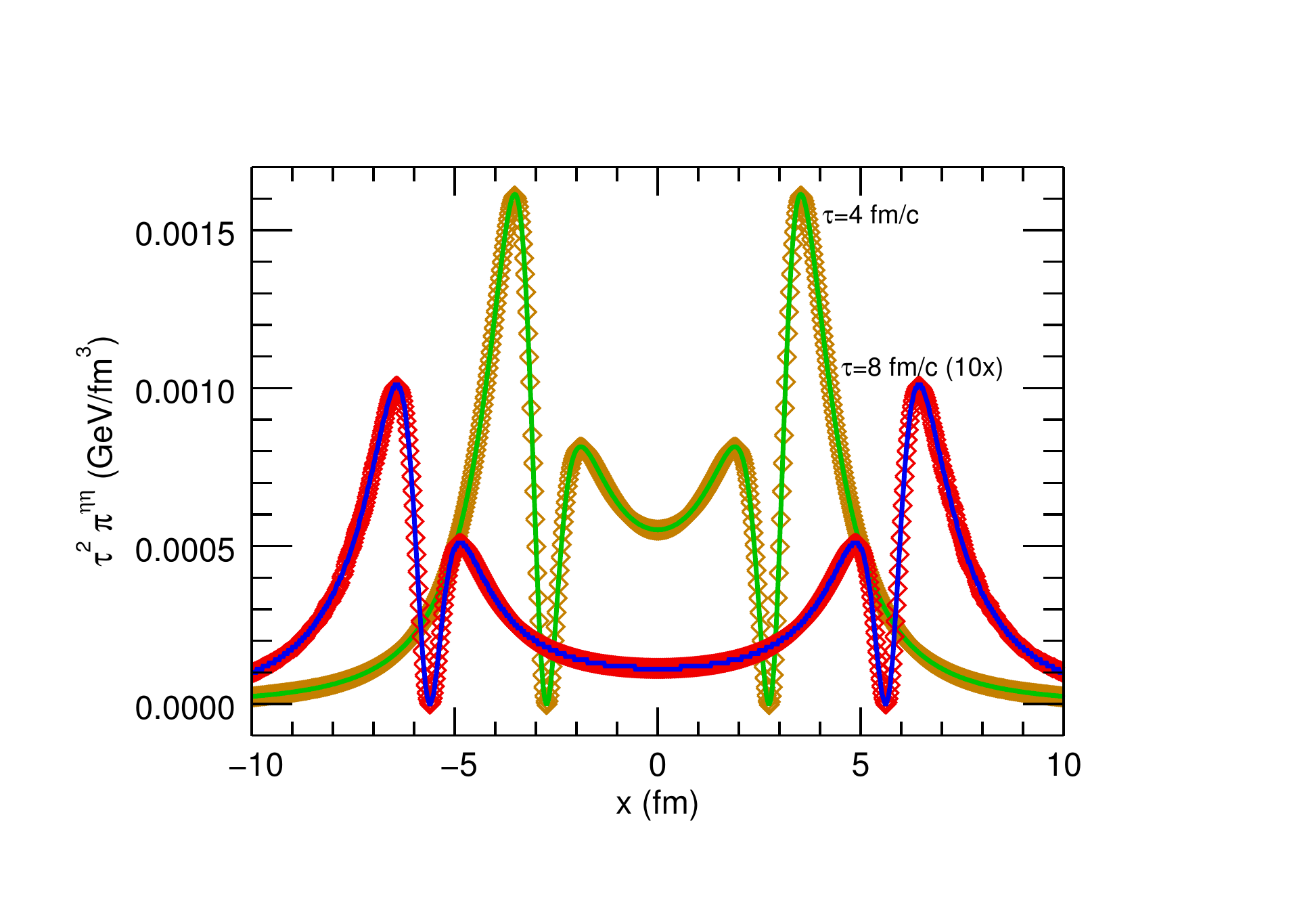}
\caption{(color online)
    Comparison between the semi-analytic solution of the Gubser viscous flow
    (solid lines) and the numerical ECHO-QGP computation (dots).}
 \label{gubserpanel}    
\end{figure*}

For the sake of clarity, we briefly summarize the main steps leading to the analytical
solution, to be compared with the numerical computation.
In the case of a conformal fluid, with $p\!=\!\epsilon/3$ EOS, the invariance for scale 
transformations sets the terms entering the second-order viscous hydrodynamic 
equations. The additional requests of azimuthal and longitudinal-boost invariance, 
constrain the solution of the hydrodynamic equations, which has to be invariant 
under ${\rm SO}(3)_q\otimes {\rm SO}(1,1)\otimes{\rm Z}_2$ transformations. 
To start with, one defines a modified space-time metric as follows (with usual 
Bjorken coordinates, $\eta$ being the spacetime rapidity):
\begin{eqnarray*}
ds^2&=&\tau^2d\hs^2\equiv\tau^2\left(\frac{d\tau^2-d\x^2}{\tau^2}-d\eta^2\right)
\nonumber \\
&=& \tau^2\left(\frac{d\tau^2-dr^2-r^2d\phi^2}{\tau^2}-d\eta^2\right),
\end{eqnarray*}
which can be viewed as a rescaling of the metric tensor:
$$
ds^2\longrightarrow d\hs^2\equiv ds^2/\tau^2\quad\Longleftrightarrow\quad
g_{\mu\nu}\longrightarrow \hat{g}_{\mu\nu}\equiv g_{\mu\nu}/\tau^2.
$$
It can be shown that $d\hs^2$ is the invariant spacetime interval of $dS_3\otimes R$, 
where $dS_3$ is the three-dimensional de Sitter space and $R$ refers to the rapidity
coordinate. It is then convenient to perform a coordinate transformation ($q$ is 
an arbitrary parameter setting an energy scale for the solution once one goes 
back to physical dimensionful coordinates)
\be\label{eq:changeofcoord}
 \sinh\rho\equiv -\frac{1-q^2(\tau^2-r^2)}{2q\tau},\quad \tan\theta\equiv
 \frac{2qr}{1+q^2(\tau^2-r^2)},
\ee
after which the rescaled spacetime element $d\hs^2$ reads
\be\label{eq:rescaledmetric}
d\hs^2=d\rho^2-\cosh^2\!\!\rho\,(d\theta^2+\sin^2\!\!\theta\, d\phi^2)-d\eta^2.
\ee
The full symmetry of the problem is now manifest. ${\rm SO}(1,1)$ and ${\rm Z}_2$ 
refer to the usual invariance for longitudinal boosts and $\eta\to -\eta$ inversion, 
while ${\rm SO}(3)_q$ reflects the spherical symmetry of the rescaled metric tensor 
in the new coordinates. In Gubser coordinates the fluid is at rest:
\be\label{eq:gubserflow}
 \hat{u}_\rho=1,\quad \hat{u}_\theta=\hat{u}_\phi=\hat{u}_\eta=0.
\ee
The corresponding flow in Minkowski space can be obtained taking into account 
both the rescaling of the metric and the change of coordinates
$$
u_\mu=\tau\frac{\partial\hx^\nu}{\partial x^\mu}\hu_\nu,
$$
where $\hx^\mu=(\rho,\theta,\phi,\eta)$ and  $x^\mu=(\tau,r,\phi,\eta)$. Other 
quantities such as the temperature or the viscous tensors require the solution 
of the following set of hydrodynamic equations (their most general form actually 
admits further terms that were derived for a system of massless particles in 
refs.~\cite{jai1,jai2}), valid for the case of a conformal fluid with $\varepsilon
=3p$:
\begin{subequations}\label{eq:gubserpanel}
\begin{align}
\frac{DT}{T}+\frac{\theta}{3}-\frac{\pi_{\mu\nu}\sigma^{\mu\nu}}{3sT}=0\label{eq:temp}\\
\tau_\pi\left(\Delta^\mu_\alpha\Delta^\nu_\beta D\pi^{\alpha\beta}+
\frac{4}{3}\pi^{\mu\nu}\theta\right)+\pi^{\mu\nu}=2\eta\sigma^{\mu\nu}\label{eq:viscous}.
\end{align}
\end{subequations}
In the case of the Gubser flow in Eq.~(\ref{eq:gubserpanel}), due to the traceless 
and transverse conditions $\hat\pi^\mu_\mu\!=\!0$ and $\hat{u}_\mu\hat\pi^\mu_\nu\!=\!0$, 
one has simply to solve the two equations ($\bar\pi^{\eta\eta}\equiv\hat\pi^{\eta\eta}/\hs\hT$)
\be\label{eq:hatT}
\frac{1}{\hT}\frac{d\hT}{d\rho}+\frac{2}{3}\tanh\rho=\frac{1}{3}\bar\pi^{\eta\eta}\tanh\rho
\ee
and ($\hat\eta/\hat{s}=\eta/s$, being the ratio dimensionless)
\be\label{eq:hattau}
\hat\tau_R\left[\frac{d\bar\pi^{\eta\eta}}{d\rho}+\frac{4}{3}\left(\bar\pi^{\eta\eta}\right)^2
\tanh\rho\right]+\bar\pi^{\eta\eta}=\frac{4}{3}\frac{\hat\eta}{\hs\hT}\tanh\rho.
\ee
The solution can be then mapped back to Minkowski space through the formulae:
\be
T=\hat{T}/\tau,\qquad\pi_{\mu\nu}=\frac{1}{\tau^2}\frac{\partial\hat{x}^\alpha}
{\partial x^\mu}\frac{\partial\hat{x}^\beta}{\partial x^\nu}\hat\pi_{\alpha\beta}.
\ee
In fig.~\ref{gubserpanel} we show the comparison between the Gubser analytical solution 
and our numerical computation for the temperature $T$ and the components $\pi^{xx}$, 
$\pi^{xy}$ and $\pi^{\eta\eta}$ of the viscous stress tensor respectively, at different 
times. The initial energy density profile was taken from the exact Gubser solution 
at the time $\tau=1$ fm/c. 
The simulation has been performed with a grid of 0.025 fm in space and 0.001 
fm in time. The shear viscosity to entropy density ratio was set to $\eta/s=0.2$, 
while the shear relaxation time is $\tau_R = 5\eta/(\varepsilon+p)$. The energy scale 
is set to $q=1$ fm$^{-1}$. As it can be seen, the agreement is excellent up to late 
times.

\subsection{T-vorticity for a viscous fluid}
\begin{figure}[!ht] 
 \includegraphics[width=0.5\textwidth]{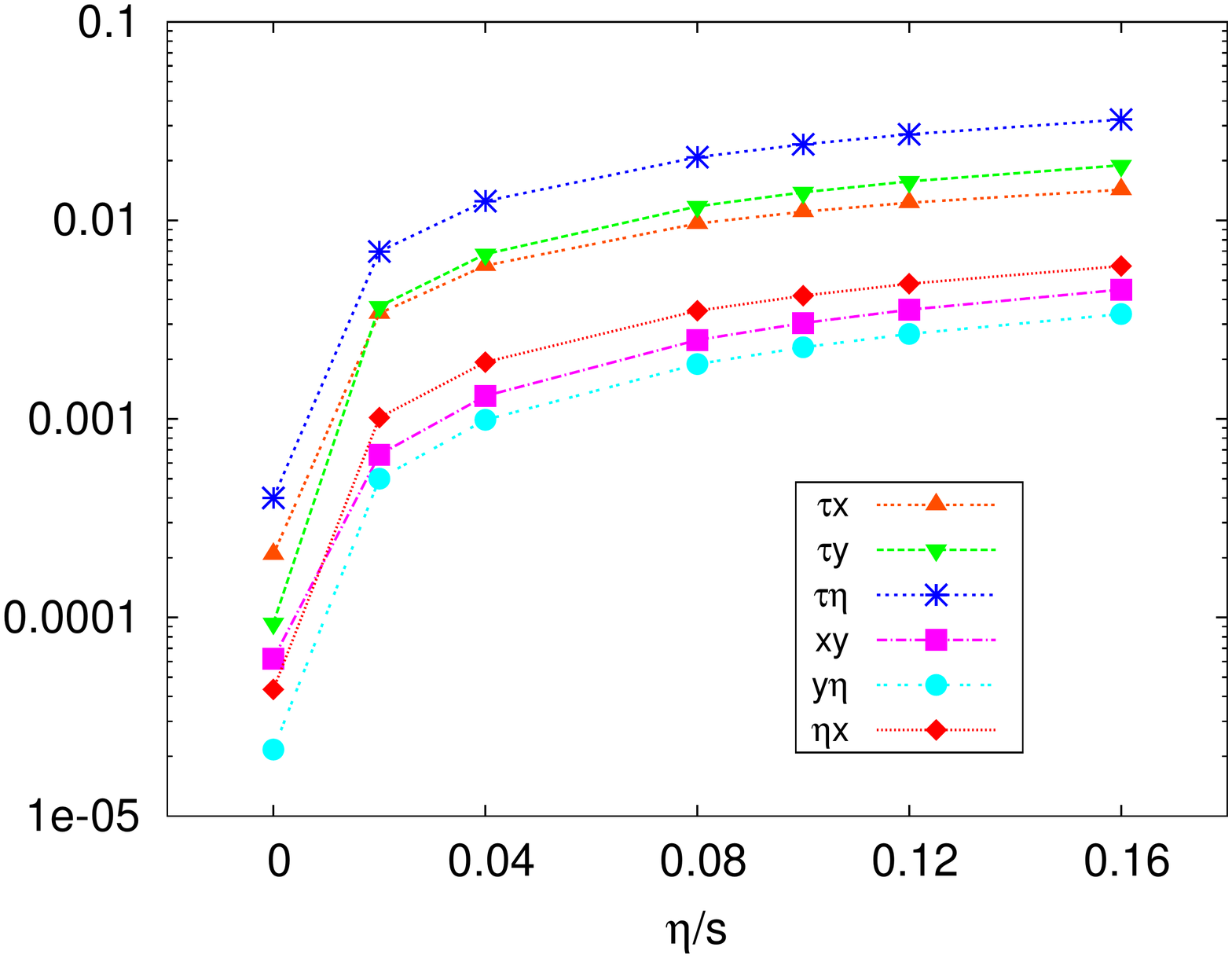}
 \hspace{-15mm}
 \includegraphics[width=0.5\textwidth]{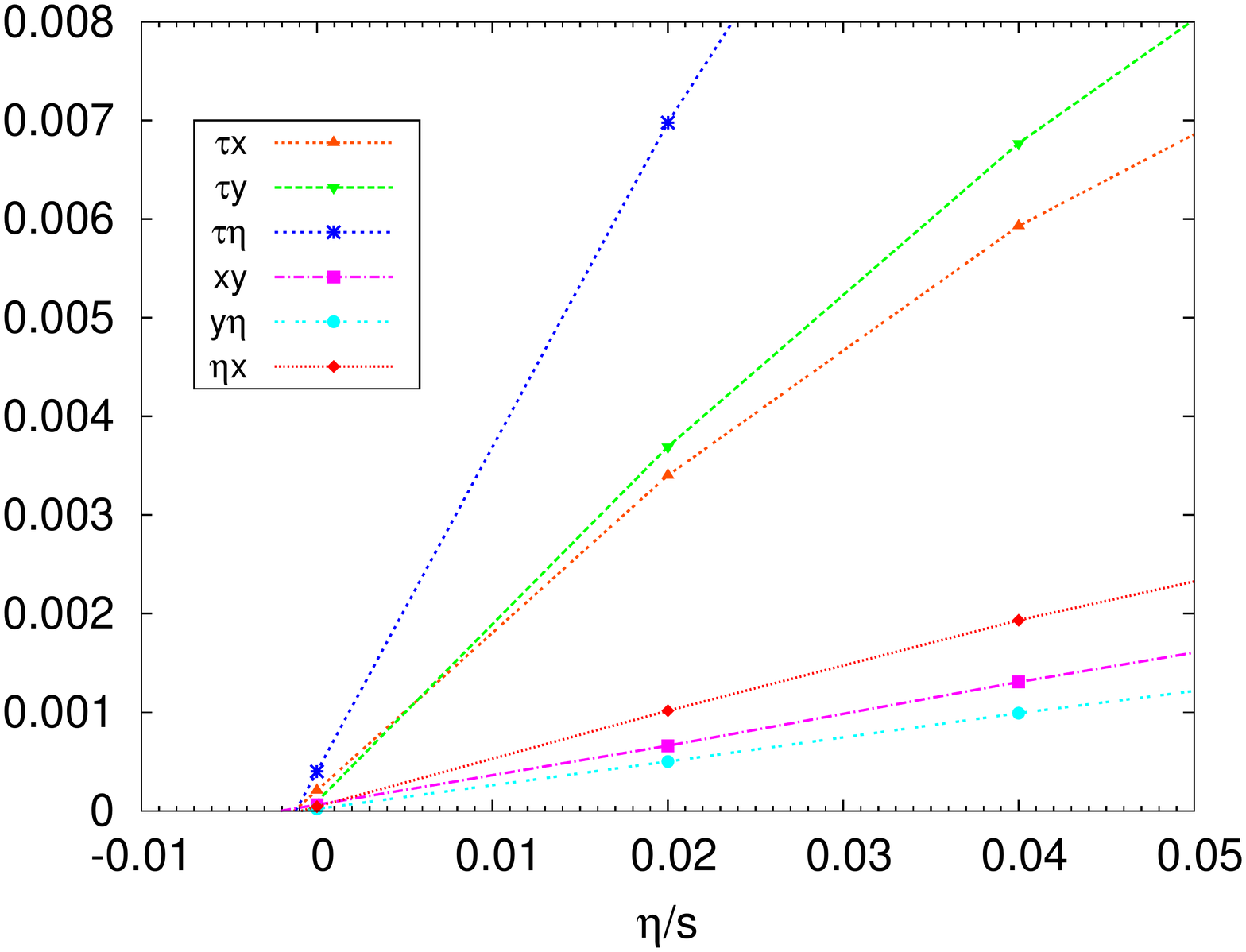}
\caption{(color online) Mean of the absolute values of $\Omega_{\mu\nu}/T^2$ components at 
the freeze-out hypersurface as a function of $\eta/s$. Note that the $\Omega_{x\eta},\Omega_
{y\eta},\Omega_{\tau\eta}$ have been multiplied by $1/\tau$. Upper panel: 
log scale. Lower panel: magnification of the region around zero viscosity.}
\label{numvisc}
\end{figure}
  
Unlike for an ideal uncharged fluid, T-vorticity can be generated in a viscous 
uncharged fluid even if it is initially vanishing. Thus, the T-vorticity can 
be used as a tool to estimate the numerical viscosity of the code in the ideal 
mode by extrapolating the viscous runs. 

A comment is in order here.
In general, in addition to standard truncation errors due to finite-difference 
interpolations, all shock-capturing upwind schemes are known to introduce numerical 
approximations that behave roughly as a dissipative effects, especially in the 
simplified solution to the Riemann problems at cell interfaces \cite{leveque}. 
It is therefore important to check whether the code is not introducing, for a 
given resolution, numerical errors which are larger than the effects induced by 
the physics. We refer to the global numerical errors generically as numerical viscosity.

We have thus calculated the T-vorticity for different physical viscosities (in fact 
$\eta/s$ ratios), in order to provide an upper bound for the numerical viscosity of 
ECHO-QGP in the ideal mode. The mean value of the T-vorticity is shown in 
fig.~\ref{numvisc} and its extrapolation to zero occurs when $|\eta/s| \lesssim 0.002$ which 
is a very satisfactory value, comparable with the one obtained in ref.~\cite{karpe}. 
The good performance is due to the use of high-order reconstruction methods that 
are able to compensate for the highly diffusive two-wave Riemann solver employed 
\cite{echoqgp}.

\section{Directed flow, angular momentum and thermal vorticity}

With the initial conditions reported at the end of the Sect.~\ref{nuclear} we have 
calculated the directed flow of pions (both charged states) at the freezeout and compared 
it with the STAR data for charged particles collected in the centrality interval 
40-80\% \cite{star08}. Directed flow is an important observable for several reasons.
Recently, it has been studied at lower energy \cite{steinh} with a hybrid fluid-
transport model (see also ref.~\cite{ivanov}). At $\snn = 200$ GeV, it has been 
calculated with an ideal 3+1D hydro code first by  Bozek~\cite{bozek}. Herein, we 
extend the calculation to the viscous regime. 
\begin{figure}[!ht]
  \includegraphics[width=0.5\textwidth]{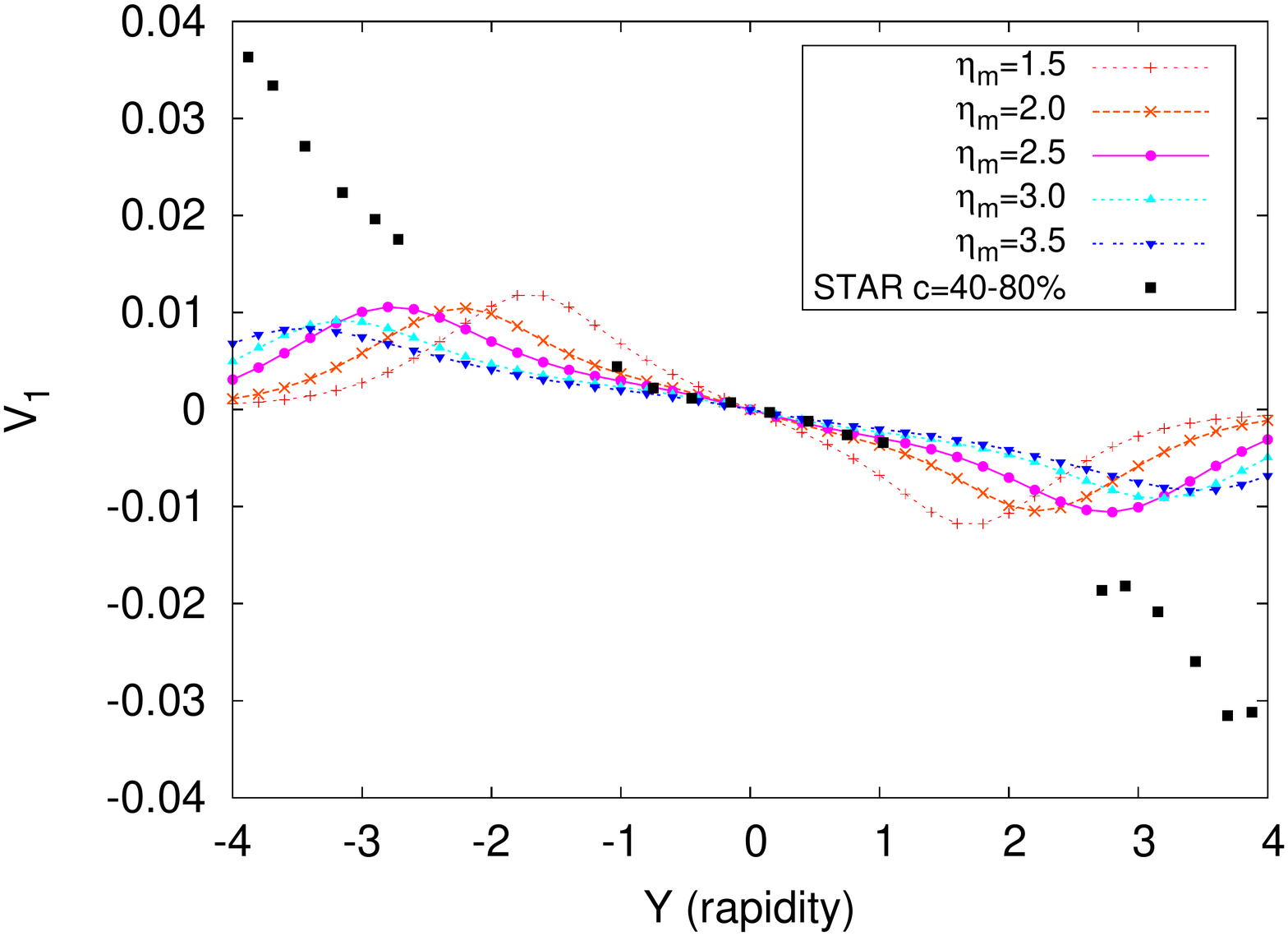}
  \caption{(color online) Directed flow of pions for different values of $\eta_m$ 
parameter with $\eta/s=0.1$ compared with STAR data \cite{star08}.}
\label{v1etam}  
\end{figure}

\begin{figure}[!ht]
\includegraphics[width=0.5\textwidth]{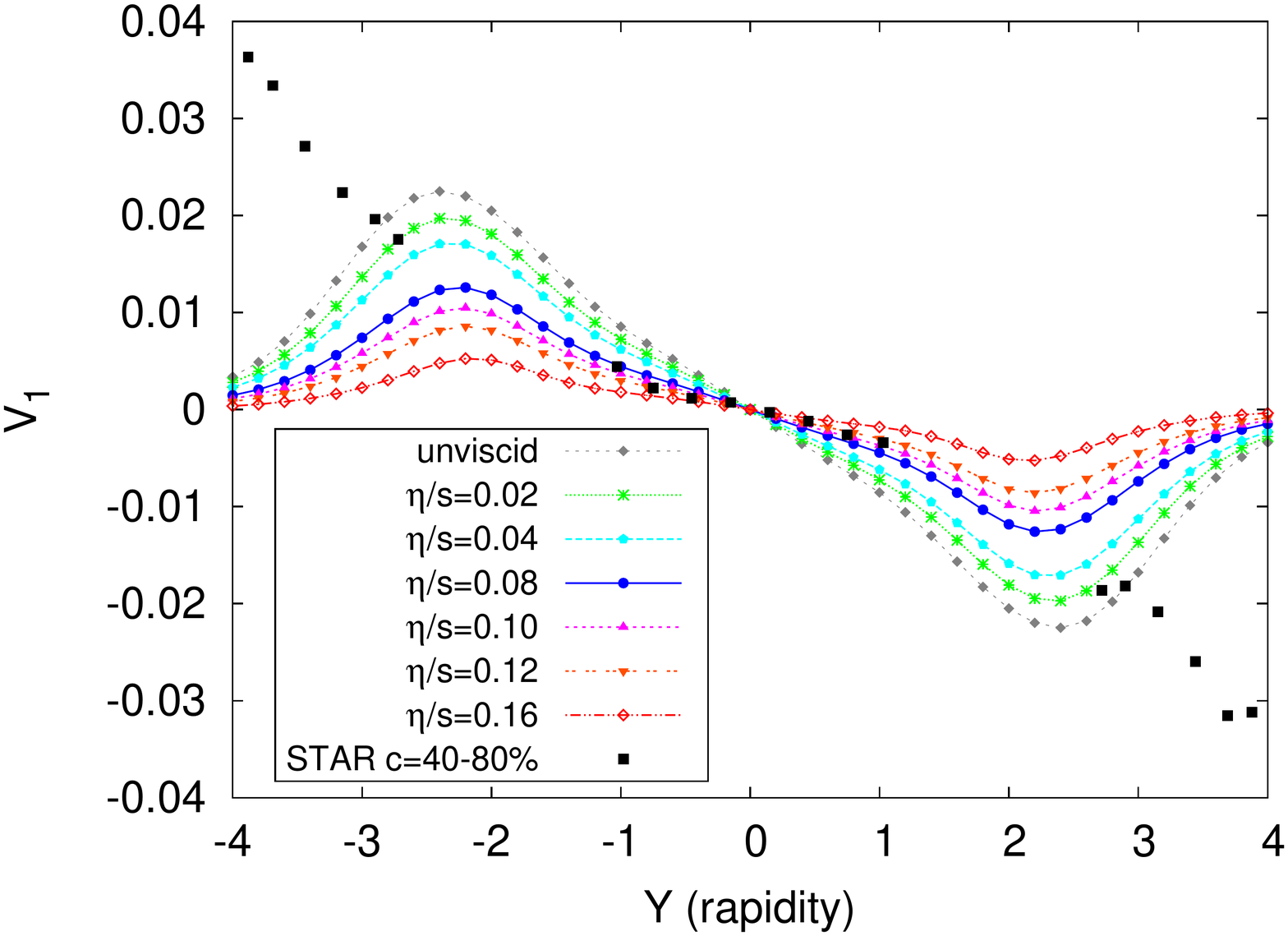}
\caption{(color online) Directed flow of pions for different values of $\eta/s$ 
with $\eta_m=2.0$ compared with STAR data \cite{star08}.}
\label{v1etas}
\end{figure}

\begin{figure}
\includegraphics[width=0.5\textwidth]{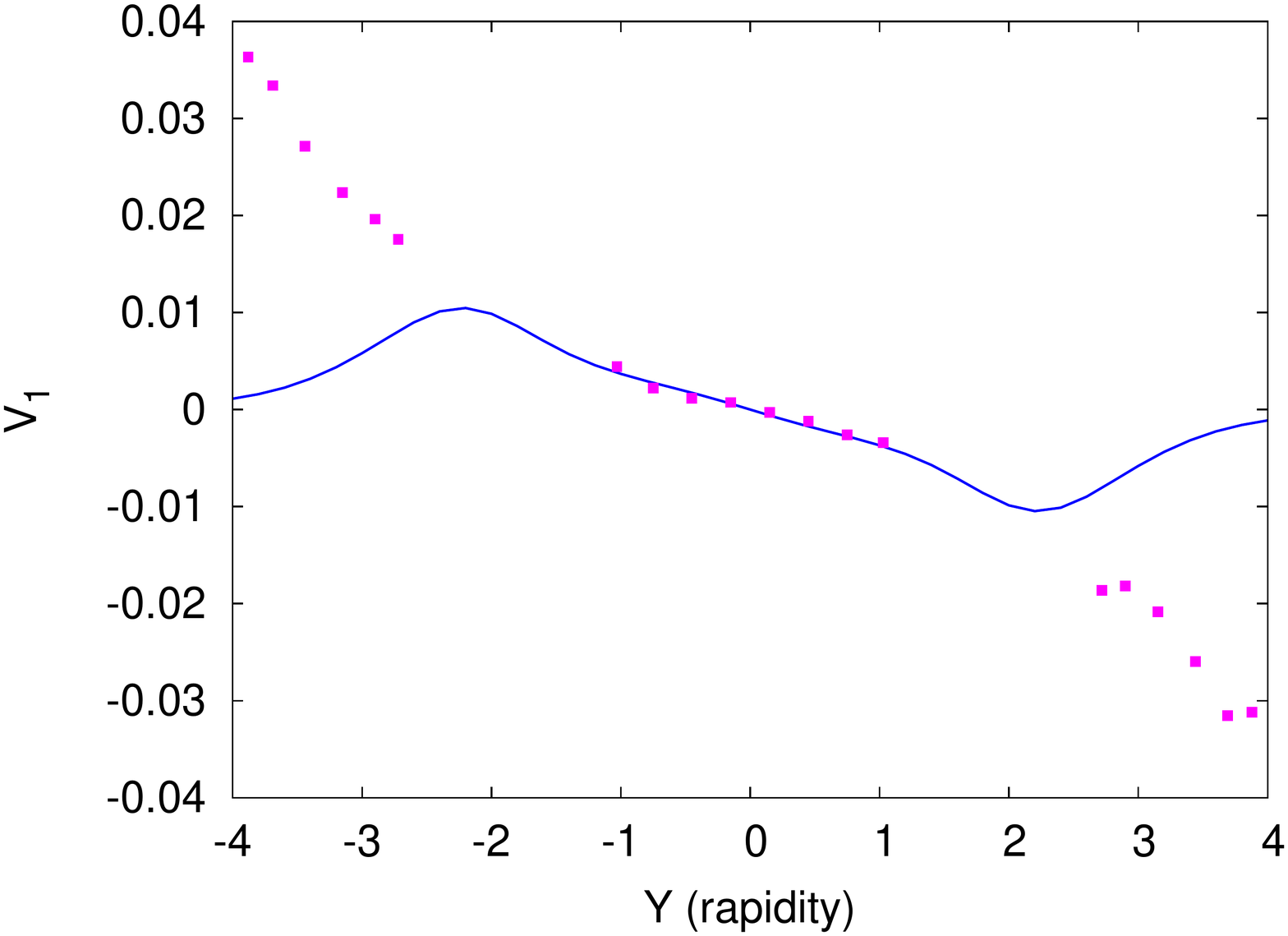}
\caption{Directed flow of pions at $\eta/s=0.1$ and $\eta_m=2.0$ compared with 
STAR data \cite{star08}.}
\label{v1final}
\end{figure}

The amount of generated directed flow at the freezeout depends of course on the initial 
conditions, particularly on the parameter $\eta_m$ (see Sect.~\ref{nuclear}), as shown 
in fig.~\ref{v1etam}. The directed flow also depends on $\eta/s$ as shown in fig.~\ref{v1etas}
and could then be used to measure the viscosity of the QCD plasma along with other 
azimuthal anisotropy coefficients. It should be pointed out that, apparently, the
directed flow can be reproduced by our hydrodynamical calculation only for $-3<y<3$.
The dependence of $v_1(y)$ on $\eta_m$ and $\eta/s$ makes it possible to adjust the 
$\eta_m$ parameter for a given $\eta/s$ value. This adjustment cannot be properly 
called a precision fit because, as we have mentioned in the Introduction, several 
effects in the comparison between data and calculations have been deliberately 
neglected in this work. However, since our aim was to obtain a somewhat realistic 
evaluation of the vorticities, we have chosen the value of $\eta_m$ for which 
we obtain the best agreement between our calculated pion $v_1(y)$ and the measured
for charged particles in the central rapidity region. For the fixed $\eta/s =0.1$ 
(approximately twice the conjectured universal lower bound) the corresponding best
value of $\eta_m$ turns out to be $2.0$ (see fig.~\ref{v1final}).

\begin{figure}[!ht]
\includegraphics[width=0.5\textwidth]{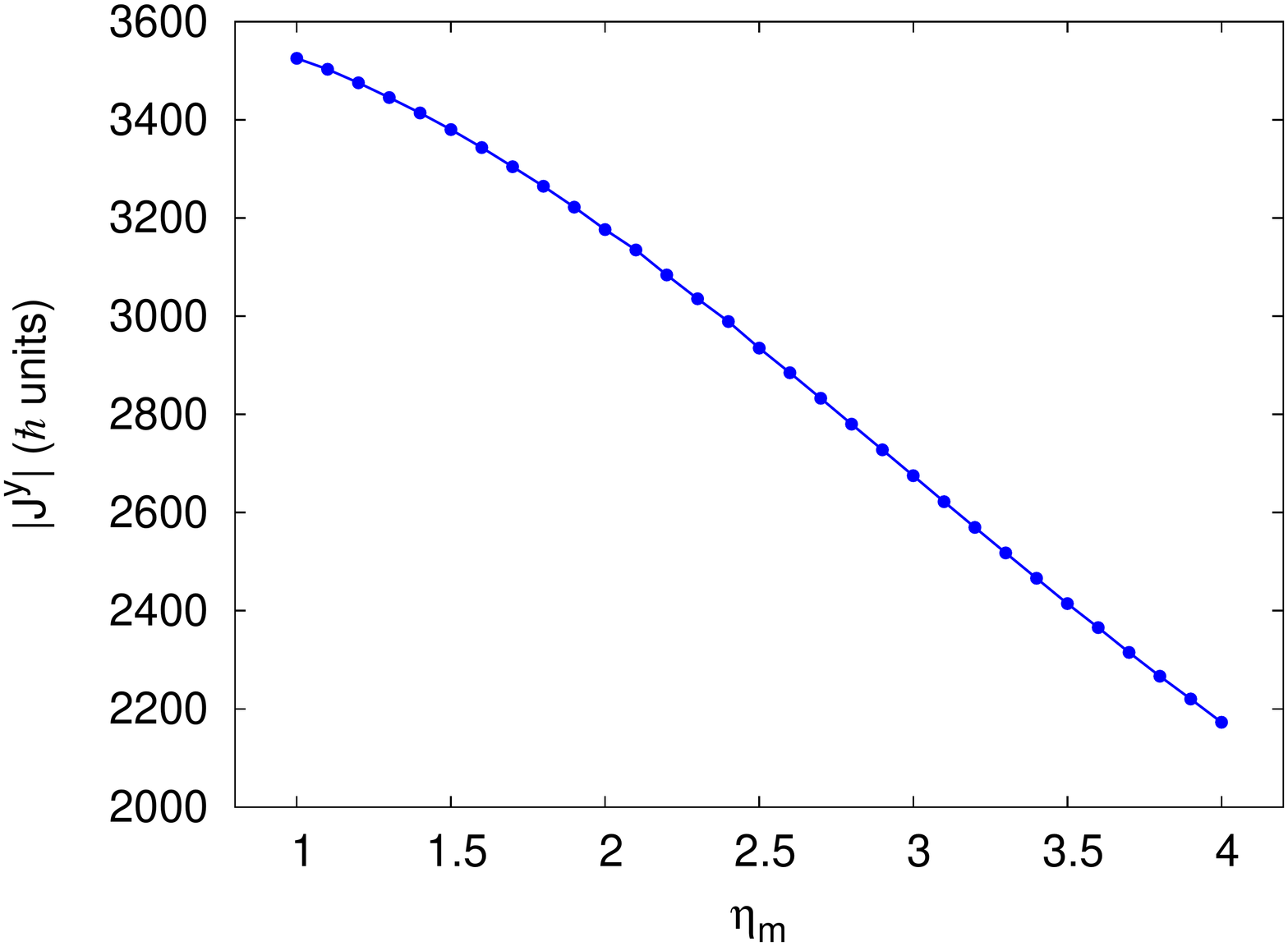}
\caption{Angular momentum (in $\hbar$ units) of the plasma with Bjorken initial
conditions as a function of the parameter $\eta_m$.}
\label{jvsetam}
\end{figure}

It is worth discussing more in detail an interesting relationship between the 
value of the parameter $\eta_m$ and that of a conserved physical quantity, the 
angular momentum of the plasma, which, for BIC is given by the integral (see Appendix 
A for the derivation):
\be\label{angmombic}
 J^y = - \tau_0 \int \di x \, \di y \, \di \eta \; x \, \varepsilon(x,y,\eta) 
 \, \sinh \eta
\ee
Since $\eta_m$ controls the asymmetry of the energy density distribution in the 
$\eta-x$ plane, one expects that $J_y$ will vary as a function of $\eta_m$.
Indeed, if the energy density profile is symmetric in $\eta$, the integral in
eq.~(\ref{angmombic}) vanishes.
Yet, for any finite $\eta_m \ne 0$, the profile (\ref{inidens}) is not symmetric and
$J_y \ne 0$ (looking at the definition of $f_+$ and $f_-$ it can be realized that 
only in the limit $\eta_m \to \infty$ the energy density profile becomes symmetric).  
The dependence of the angular momentum on $\eta_m$ with all the initial parameters
kept fixed is shown in fig.~\ref{jvsetam}. For the value $\eta_m =2.0$ it turns out
to be around $3.18 \times 10^3$ in $\hbar$ units.

\begin{figure}[!ht]
\includegraphics[width=0.5\textwidth]{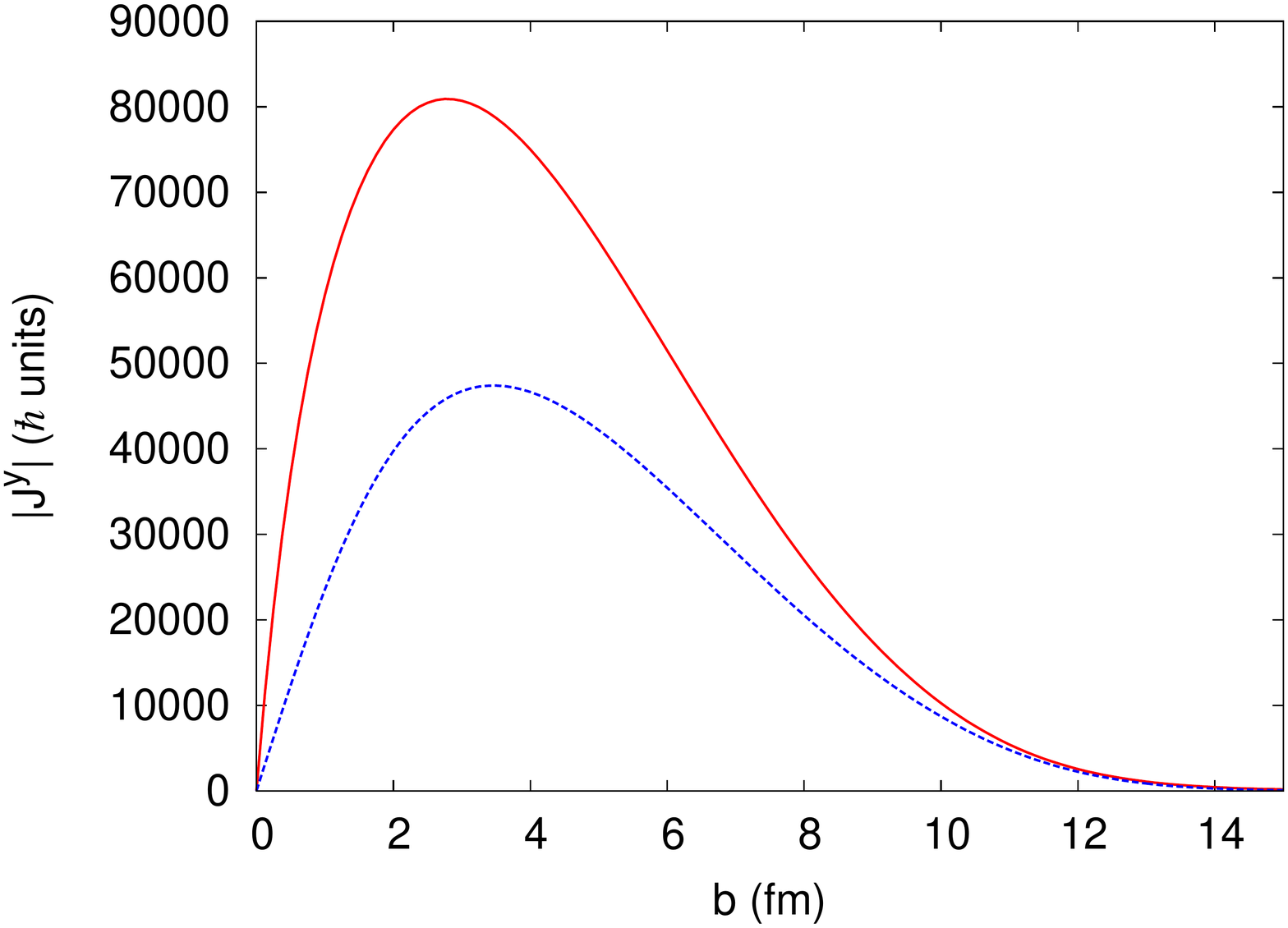}
\caption{(color online) Estimated angular momentum (in $\hbar$ units) of the 
overlap region of the two colliding nuclei (solid line) and total angular 
momentum of the plasma according to the parametrization of the initial conditions
(dashed line), as a function of the impact parameter.}
\label{jvsb}
\end{figure}

It is also interesting to estimate an upper bound on the angular momentum of the plasma by 
evaluating the angular momentum of the overlap region of the two colliding nuclei. This 
can be done by trying to extend the simple formula for two sharp spheres. In our 
conventional reference frame, the initial angular momentum of the nuclear overlap 
region is directed along the $y$ axis with negative value and can be written as:
\be
 J^y = \int \di x \,\di y \; w(x,y) (T_+ - T_-) x \dfrac{\sqrt{s_{NN}}}{2}
\ee
where $T_\pm$ are the thickness functions like in eq.~\ref{thick} and
$$
 w(x,y)= \frac{\min(n(x+b/2,y,0),n(x-b/2,y,0))}{\max(n(x+b/2,y,0),n(x-b/2,y,0))}
$$
is the function which extends the simple product of two $\theta$ functions used for the
overlap of two sharp spheres. Note that the $\w(x,y)$ is 1 for full overlap ($b$=0) and
implies a vanishing angular momentum for very large $b$ (see fig.~\ref{jvsb}) (see 
also ref.~\cite{vovchenko}). 

At $b=11.57$ fm the above angular momentum is about $3.58 \times 10^3$ in $\hbar$ units. 
This means that, with the current parametrization of the initial conditions, for
that impact parameter about 89\% of the angular momentum is retained by the hydrodynamical 
plasma while the rest is possibly taken away by the corona particles.

\begin{figure}
\includegraphics[width=0.5\textwidth]{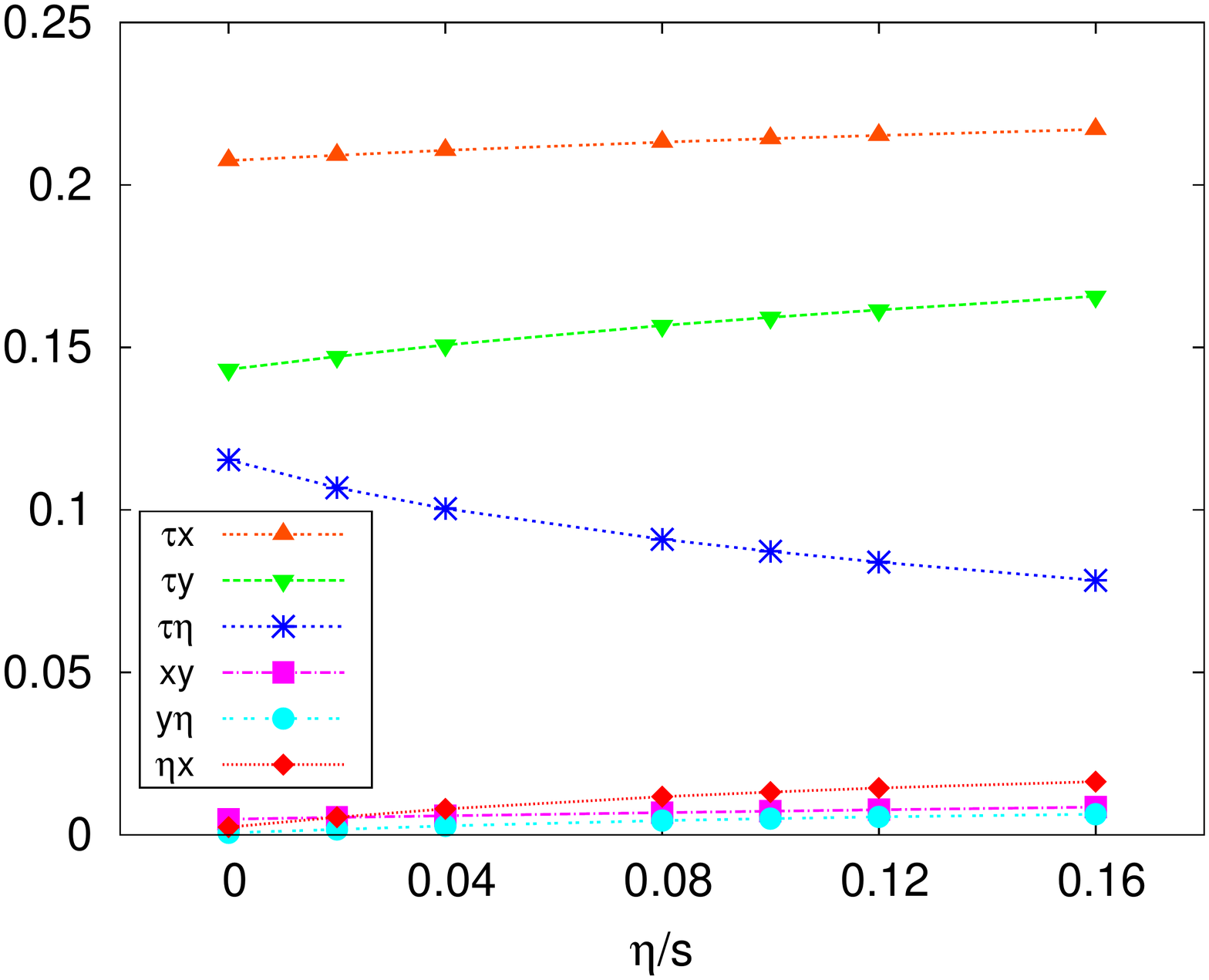}
\caption{(color online) Mean of the absolute value of thermal vorticity covariant 
components at the freeze-out as a function of $\eta/s$. Note that the 
$\thv_{x\eta},\thv_{y\eta}, \thv_{\tau\eta}$ have been multiplied by $1/\tau$.}
\label{thvort1}
\end{figure}

\begin{figure}
\includegraphics[width=0.5\textwidth]{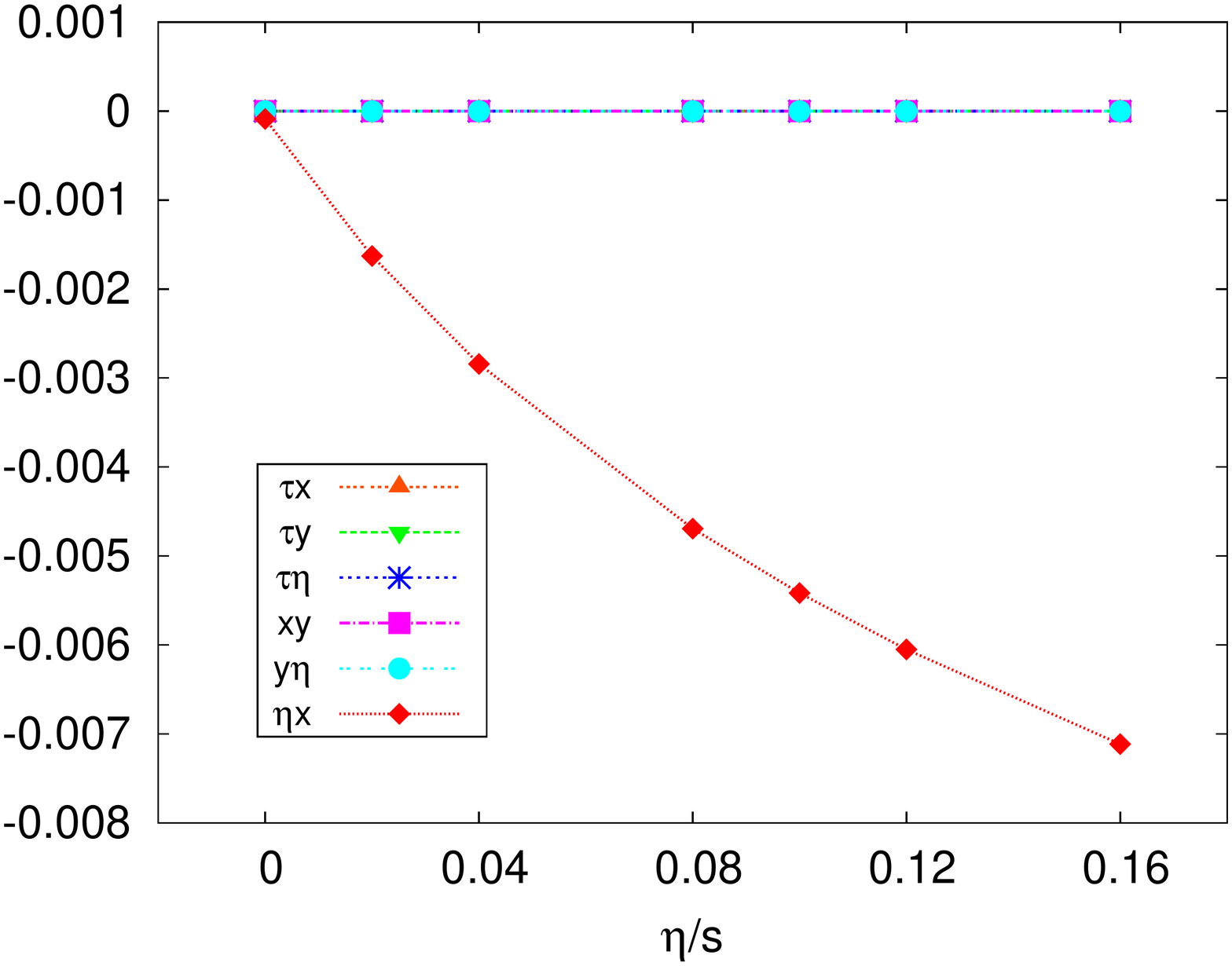}
\caption{(color online) Mean values of thermal vorticity components at the freeze-out 
as a function of $\eta/s$. Note that the $\thv_{x\eta},\thv_{y\eta},\thv_{\tau\eta}$ 
have been multiplied by $1/\tau$.}
\label{thvort2}
\end{figure}

\begin{figure}
\includegraphics[width=0.5\textwidth]{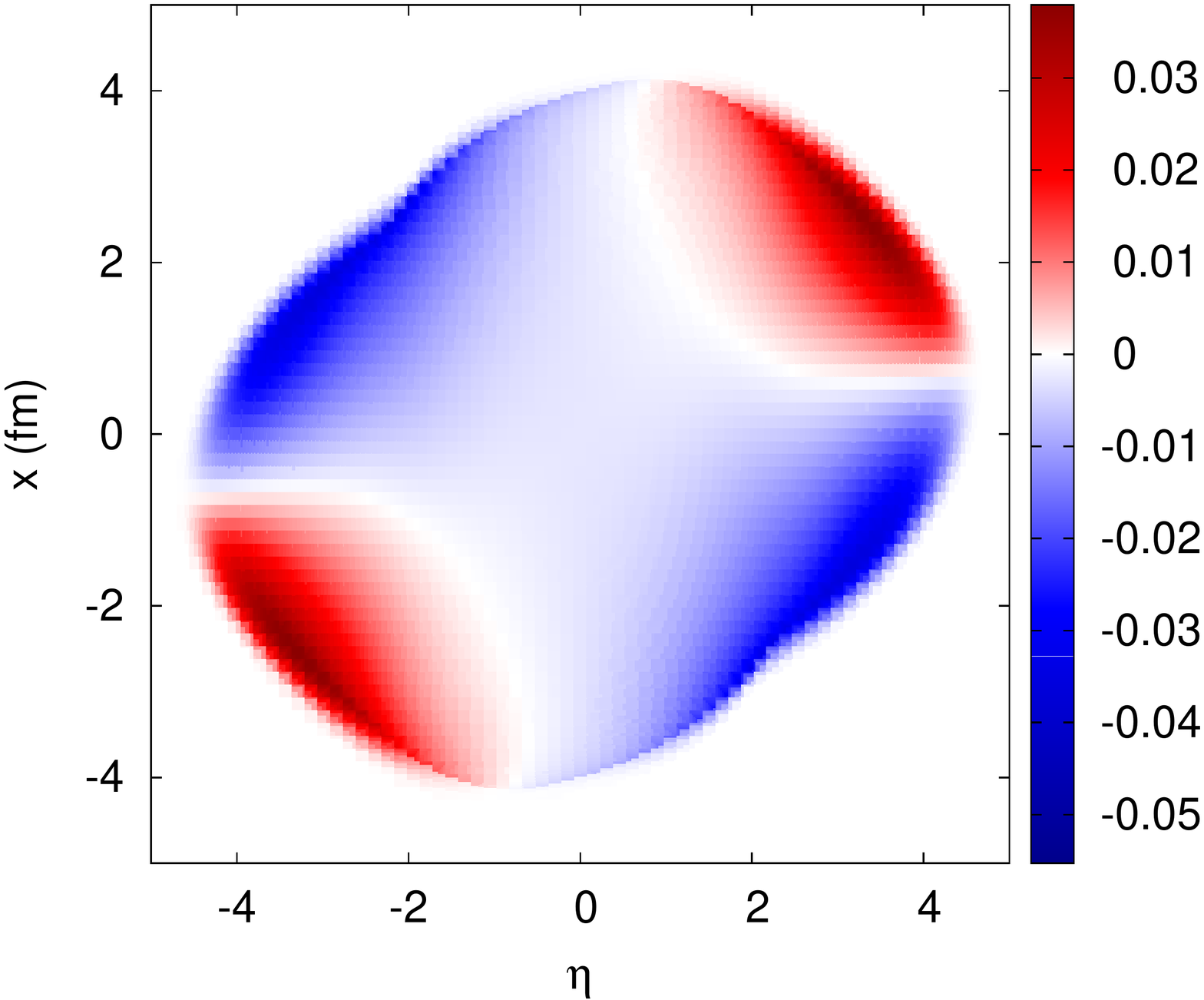}
\caption{(color online) Contour plot of $1/\tau$-scaled $\eta x$ covariant component 
of the thermal vorticity, $\thv_{\eta x}/\tau$ over the freeze-out hypersurface for 
$y=0$, $\eta/s$=0.1, $\eta_m$=2.0.}
\label{thvmap}
\end{figure}

With the final set of parameters, we have calculated the thermal vorticity $\thv$.
As it has been mentioned in Sect.~\ref{defin}, this vorticity is adimensional
in cartesian coordinates) and it is constant at global thermodynamical equilibrium 
\cite{becacov}, e.g. for a globally rotating fluid with a rigid velocity field. 
In relativistic nuclear collisions we are far from such a situation, nevertheless 
some thermal vorticity can be generated, both in the ideal and viscous case. This 
is shown in figs.~\ref{thvort1} and \ref{thvort2}.
It can be seen that the generated amount of thermal vorticity has some non-trivial
dependence on the viscosity. Particularly, as it is apparent from fig.~\ref{thvort2}, the 
$\thv_{x\eta}$ component - which is directed along the initial angular momentum -
has a non-vanishing mean value whose magnitude significantly increases with increasing 
viscosity. Its map at the freezeout, for a fixed value of the $y$ coordinate $y=0$, is 
shown in fig.~\ref{thvmap} where it can be seen that
it attains a top (negative) value of about 0.05 corresponding to a kinematical vorticity,
at the freezeout temperature of 130 MeV, of about 0.033 $c$/fm $\simeq 10^{22} {\rm s}^{-1}$. 
In this respect, the Quark Gluon Plasma would be the fluid with the highest vorticity 
ever made in a terrestrial laboratory. However, the mean value of this component at 
the same value of $\eta/s = 0.1$ is of the order of $5.4 \times 10^{-3}$, that is about 
ten times less than its peak value, as shown in fig.~\ref{thvort2}. This mean thermal
vorticity is the consistently lower than the one estimated in ref.~\cite{beca2} 
(about 0.05) with the model described in refs.~\cite{csernai1,csernai2} implying 
an initial non-vanishing transverse kinematical and thermal vorticity $\thv^\Delta$. 
This reflects in a quite low value of the polarization of $\Lambda$ baryons, as it 
will be shown in the next section.
\begin{figure*}[!ht]
    \subfloat[\label{fig:pi0_t}]{%
      \includegraphics[width=0.45\textwidth]{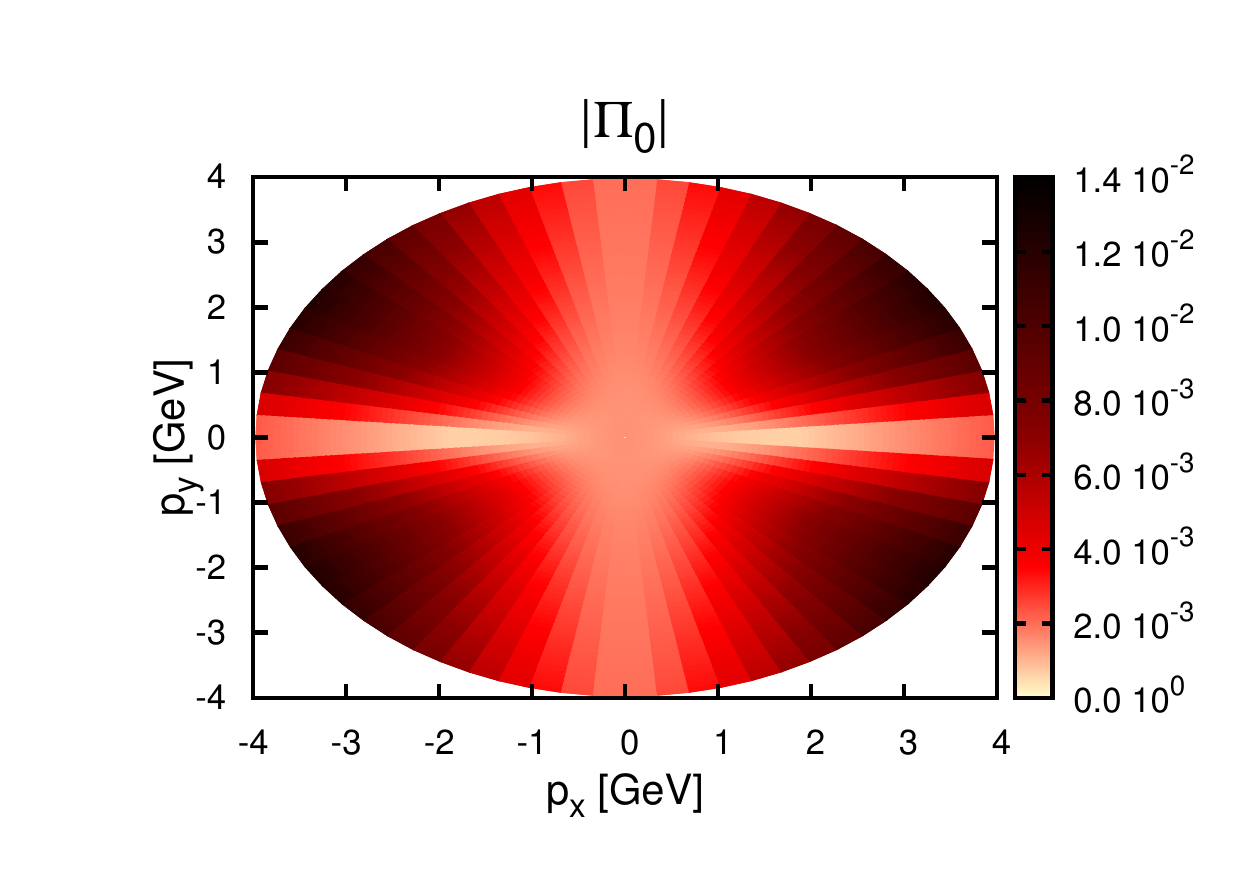}
    }
    \hfill
    \subfloat[\label{fig:pi0_x}]{%
      \includegraphics[width=0.45\textwidth]{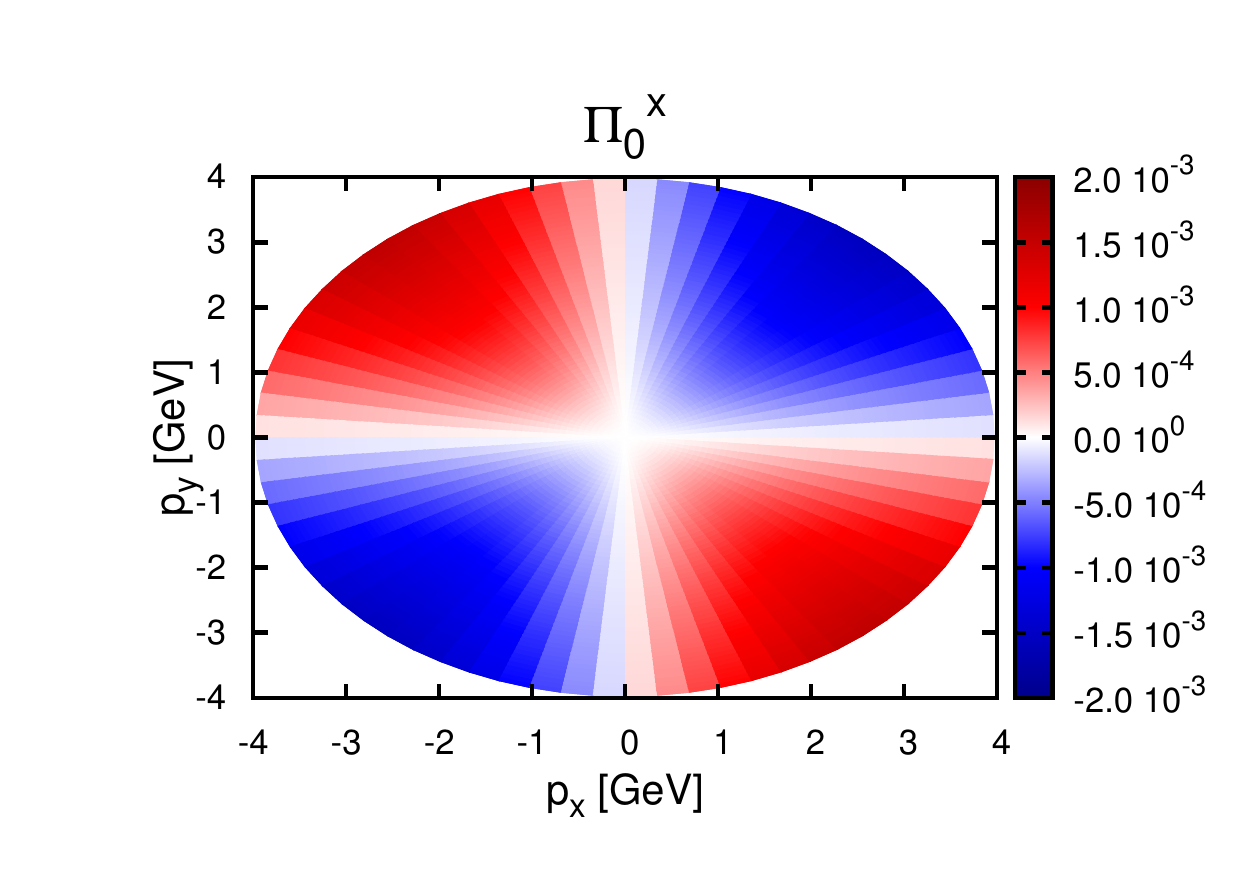}
    }\\
    \subfloat[\label{fig:pi0_y}]{%
      \includegraphics[width=0.45\textwidth]{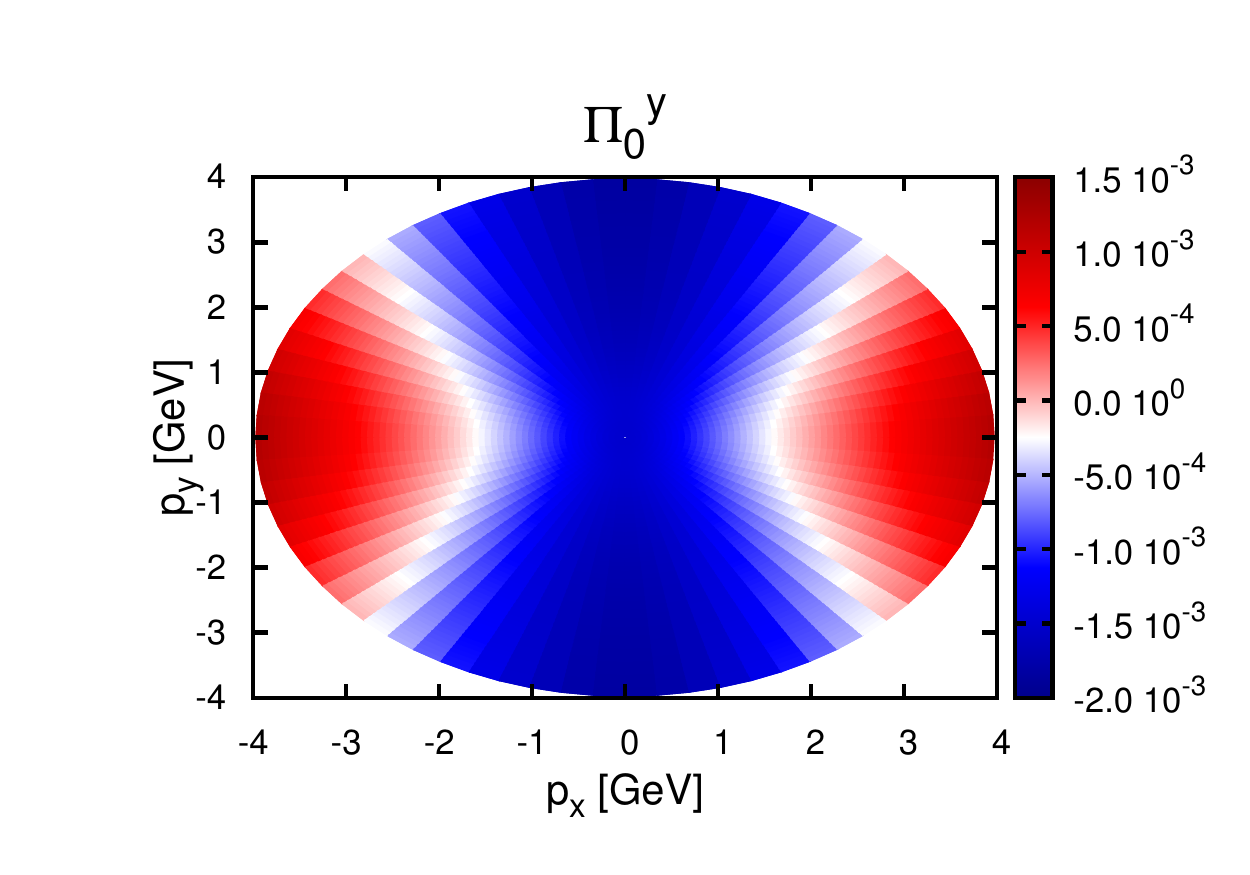}
    }
    \hfill
    \subfloat[\label{fig:pi0_z}]{%
      \includegraphics[width=0.45\textwidth]{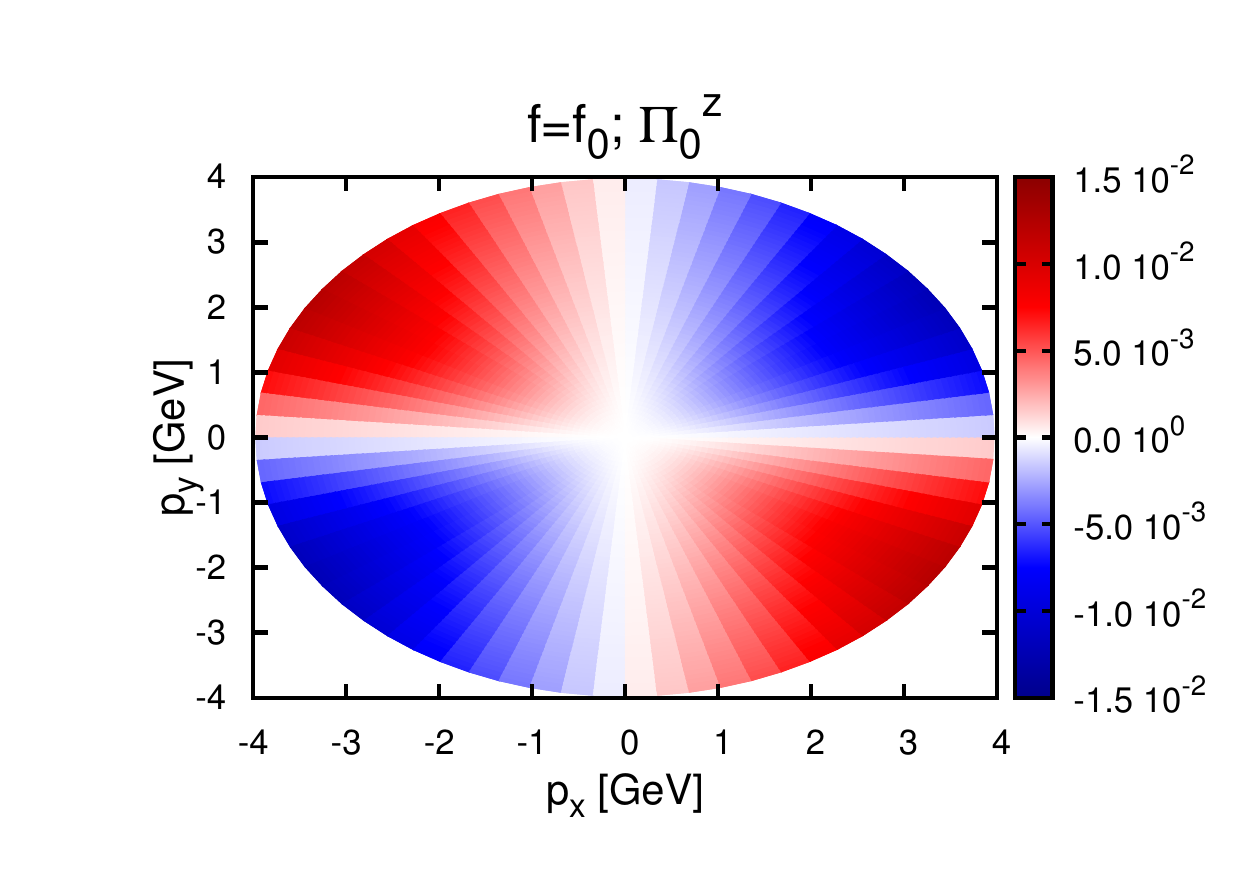}
    }\\
    \caption{(color online)
    Magnitude (panel a) and components (panels b,c,d) of the polarization vector 
     of the $\Lambda$ hyperon in its rest frame.}
    \label{polarization}
  \end{figure*}

\section{Polarization}

As it has been mentioned in the Introduction, vorticity can result in the polarization
of particles in the final state. The relation between the polarization vector of a
spin $1/2$ particle and thermal vorticity in a relativistic fluid was derived in 
ref.~\cite{beca1} and reads:
\bea\label{polint}
 \Pi^\mu(p)&=\dfrac{1}{8m} \dfrac{\int_\Sigma \di \Sigma_\lambda p^\lambda
   n_F (1-n_F) \: p_\sigma \epsilon^{\mu\nu\rho\sigma} \partial_\nu \beta_\rho}
 {\int_\Sigma \mathrm{d}\Sigma_\lambda p^\lambda \, n_F}
\eea 
where $n_F$ is the Fermi-Dirac-Juttner distribution function (\ref{juttner}) and the 
integration is over the freeze-out hypersurface $\Sigma$. The interesting feature of
this relation is that it makes it possible to obtain an indirect measurement of the 
mean thermal vorticity at the freezeout by measuring the polarization of some hadron.
For instance, the polarization of $\Lambda$ baryons, as it is well known, can be determined
with the analysis of the angular distribution of its decay products, because of parity 
violation. The polarization pattern depends on the momentum of the decaying particle, 
as it is apparent from eq.~(\ref{polint}). 

The formula (\ref{polint}) makes sense only if the components of the integrand are 
Minkowskian, as an integrated vector field yields a vector only if the tangent spaces 
are the same at each point. Before summing over the freezeout hypersurface we have 
then transformed the components of the thermal vorticity from Bjorken coordinates to 
Minkowskian by using the known rules. The thus obtained polarization vector $\Pi(p)$ 
is the one in the collision frame. However, the polarization vector which is measurable 
is the one in the decaying particle rest frame which can be obtained by means of the 
Lorentz transformations:
\bea\label{eq:polarization_0}
 \Pi_0^0&=\dfrac{\epsilon}{m}\Pi^0-\dfrac{\mathbf{p}\cdot \mathbf{\Pi}}{m}\nonumber\\
 \mathbf{\Pi_0}&=\mathbf{\Pi}-\dfrac{\mathbf{p}\cdot \mathbf{\Pi}}
 {\epsilon(m+\epsilon)}\mathbf{p}
\eea

In figure \ref{polarization} we show the $\Lambda$ polarization vector components, as 
well as its modulus, as a function of the transverse momentum ${\bf p}_T$ for $p_z=0$ 
expected under the assumptions of local thermodynamical equilibrium for the spin degrees 
of freedom maintained till kinetic freezeout. It can be seen that the polarization vector 
has quite an assorted pattern, with an overall magnitude (see fig.~\ref{polarization}, panel
(a)) hardly exceeding 1\% at momenta around 4 GeV. As expected, the $y$ component is 
predominantly negative, oriented along the initial angular momentum vector and a 
magnitude of the order of 0.1\%. Indeed, the main contribution to the polarization 
stems from the longitudinal component $\Pi_0^z$, with a maximum and minumum along 
the bisector $|p^x|=|p^y|$.

The obtained polarization values are - as expected - consistently smaller than those 
estimated in ref.~\cite{beca2} (of the order of several percent with a top value of
8-9\%) with the already mentioned initial conditions used in refs.~\cite{csernai1,
csernai2}. This is a consequence of the much lower value of the implied thermal 
vorticity, as discussed in the previous section. Also, the $\Pi_0^y$ pattern is 
remarkably different, with different location of maxima and minima.

\section{Conclusions, discussion and outlook}
\label{conclu}

To summarize, we have calculated the vorticities developed in peripheral ($b=11.6$ fm)  
nuclear collisions at $\snn =200$ GeV ($b=11.6$ fm) with the most commonly used initial 
conditions in the Bjorken hydrodynamical scheme, by using the code ECHO-QGP implementing 
second-order, causal, relativistic dissipative hydrodynamics. An extensive testing of 
the high accuracy and very low numerical diffusion properties of the code has been
carried out, followed by long-time simulations (up to $\tau=8$~fm/c) of the so-called 
viscous Gubser flow, a stringent test of numerical implementations of Israel-Stewart 
theory in Bjorken coordinates.

We have found that the magnitude of the $1/\tau$ $x-\eta$ component of the thermal vorticity 
at freezeout can be as large as $5 \times 10^{-2}$ and yet its mean value is not large 
enough to produce a polarization of $\Lambda$ hyperons much larger than 1\%, which is 
a consistently lower estimate in comparison with other recent calculations based on 
different initial conditions. We have found that the magnitude of directed flow, at 
this energy, has an interestingly sizeable dependence on both the shear viscosity 
and the longitudinal energy density profile asymmetry parameter $\eta_m$ which in 
turn governs the amount of initial angular momentum retained by the plasma. 

\begin{figure}[!ht]
\includegraphics[width=0.5\textwidth]{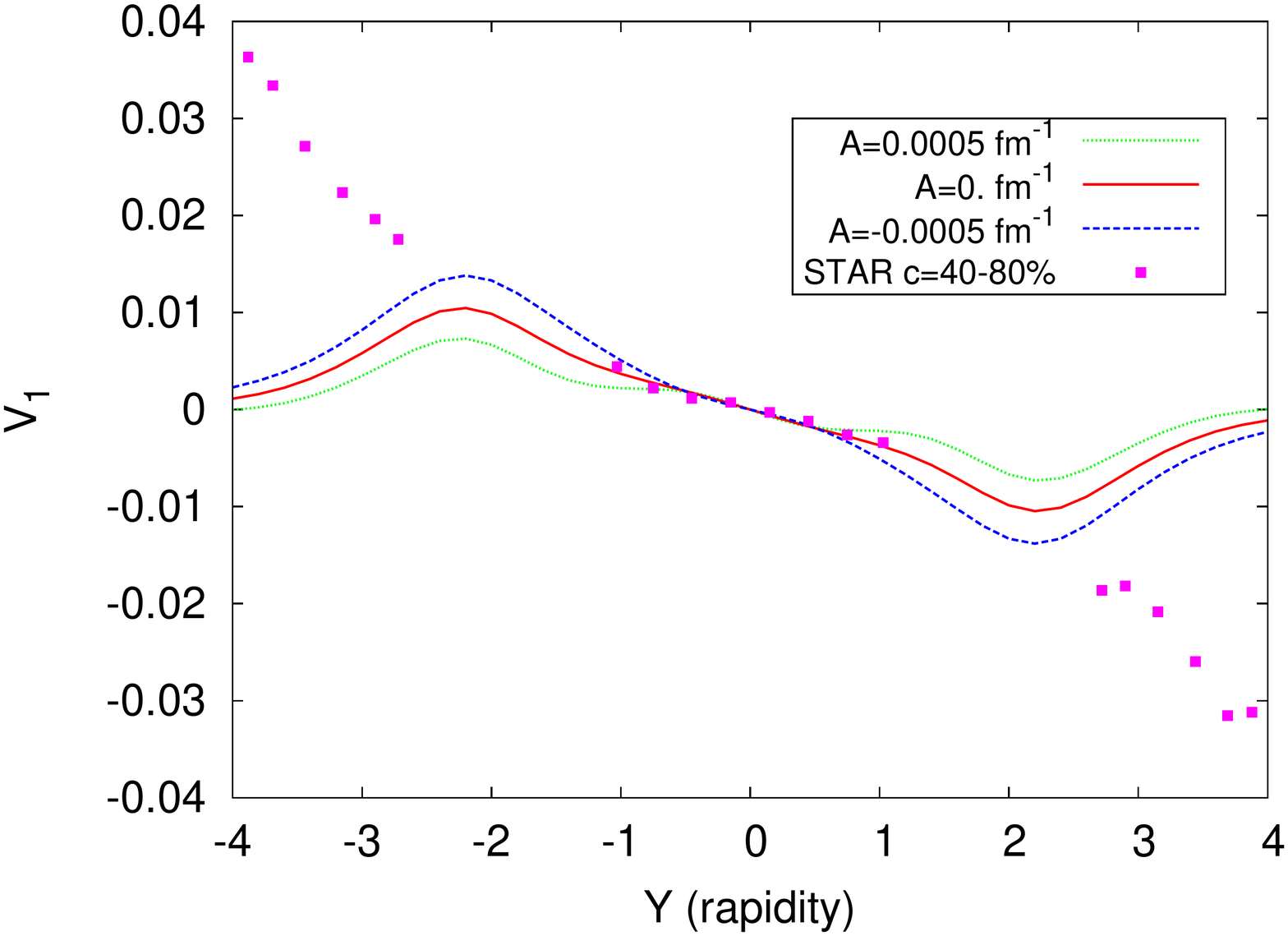}
\caption{(color online) Directed flow of pions at $\eta/s=0.1$ and $\eta_m=2.0$ 
and with the initial $u^\eta$ in the eq.~(\ref{ueta}) compared with STAR data \cite{star08}.}
\label{v1a}
\end{figure}

The fact that in 3+1D the plasma needs to have an initial angular momentum in order to
reproduce the observed directed flow raises the question whether the Bjorken initial 
condition $u^\eta = 0$ is a compelling one or, instead, the same angular momentum can be
obtained with a non trivial $u^\eta$ and with a suitable change of the energy density
profile. For a testing purpose, we have run ECHO-QGP with an initial profile:
\be\label{ueta}
  u^\eta = \frac{1}{\tau} \tanh Ax \; \sinh (y_{\rm beam} -|\eta|) 
\ee
which meets the causality constraint (see Appendix B). It is found that the directed
flow is very sensitive to an initial $u^\eta$. For a small positive value of the 
parameter $A = 5 \times 10^{-4}$ fm$^{-1}$ corresponding to a $J_y = 3.32 \times 10^3$, keeping
all other parameters fixed, the directed flow exhibits two slight wiggles around midrapidity 
(see fig.~\ref{v1a}) which are not seen in the data. For a very small negative value
of the parameter $A = -5 \times 10^{-4}$ fm$^{-1}$, corresponding to $J_y = 3.08 \times 10^3$, 
the directed flow increases while approximately keeping the same shape as for $A=0$ 
around midrapidity. However, more detailed studies are needed to determine whether
a non-vanishing initial flow velocity is compatible with the experimental observables.

We plan to extend this kind of calculation to different centralities, different 
energies and with initial state fluctuations in order to determine the possibly best
conditions for vorticity formation in relativistic nuclear collisions.

\section*{Acknowledgments}

We are grateful to P. Bozek, L. Csernai and Y. Karpenko for very useful comments
and suggestions.



\section*{APPENDIX A - Angular momentum}

The calculation of the total angular momentum of the plasma can be done provided that 
initial conditions are such that energy density falls off rapidly at large $|\eta|$.
This condition, which is met by the profile in eq.~(\ref{inidens}), indeed implies 
that a boundary exists where the angular momentum density tensor (that is the integrand 
below) vanishes and the following integral is conserved:
\be\label{angmom}
  J^{\mu\nu} = \int_\Sigma \di \Sigma_\lambda \; (x^{\mu} T^{\lambda\nu} - 
  x^{\nu} T^{\lambda\mu})
\ee
where $\Sigma$ is {\em any} spacelike hypersurface extending over the region where
the angular momentum density vanishes. The obvious choice for Bjorken-type initial 
conditions is the hypersurface $\tau=\tau_0$. 

It should be stressed that a vector (or tensor) integral is meaningful in flat spacetime
only if the components are the cartesian ones. Hence, for the hypersurface $\tau=\tau_0$,
the integration variables are conveniently chosen to be the Bjorken ones, but the 
components of the stress-energy tensor as well as the $x$ vector will be cartesian.
Since the only non vanishing component of the angular momentum in our conventional
reference frame is $J^y$, orthogonal to the reaction plane, we can write:
\be\label{angmom2}
 J^y = J^{31} = \int_\Sigma \di \Sigma_\lambda \; \left[x^3 T^{\lambda 1} - 
  x^{1} T^{\lambda 3}\right].
\ee
Finding the hypersurface measure $\di \Sigma_\lambda$ in cartesian components, but expressed through
Bjorken variable, requires some 
reasoning. First, one has to remind that:
\be\label{meas1}
 \di \Sigma_\lambda = \di \Sigma \, n_\lambda
\ee
where $n$ is the unit vector normal to the hypersurface $\tau=\tau_0$ which is readily
found to be (cartesian covariant components): 
\be\label{meas2}
  n_\mu = (\cosh \eta, 0, 0, - \sinh \eta)
\ee
Now, since Bjorken coordinates are time-orthogonal ($g_{\tau i}=0$) and with $g_{\tau\tau} = 1$, 
the invariant spacetime measure $\di \Omega$ can be factorized into the product of the
infinitesimal ``time'' $\di \tau$ and the infinitesimal measure of the orthogonal hypersurface
$\di \Sigma$:
$$
 \di \Omega = \di \tau \, \di \Sigma
$$
At the same time:
$$
 \di \Omega = \sqrt{|g|} \di \tau \, \di x \, \di y \, \di \eta = \tau \, \di \tau 
 \, \di x \, \di y \, \di \eta  
$$
whence:
\be\label{meas3}
   \di \Sigma = \tau \, \di x \, \di y \, \di \eta
\ee
Using eqs.~(\ref{meas1}), (\ref{meas2}) and (\ref{meas3}), eq.~(\ref{angmom2}) can
be written as:
\bea\label{angmom3}
 J^y = && \tau \int \di x \, \di y \, \di \eta\; \left[ \cosh \eta (x^3 T^{0 1} 
 - x^{1} T^{0 3}) \right. \nonumber \\
 && \left. - \sinh \eta (x^3 T^{3 1} - x^{1} T^{3 3}) \right]
\eea
At the time $\tau=\tau_0$, the stress-energy tensor is supposedly the ideal one and 
there is no transverse velocity, so that $T^{01}=T^{31}=0$, while $T^{33} = (\varepsilon + p) 
u^{z}u^z + p$ and $T^{03} = (\varepsilon+p) u^\0 u^z$. Pluggin these expressions into the
(\ref{angmom3}) along with the transformation equation:
\be\label{transf1}
  t = \tau \cosh \eta \qquad x=x \qquad y=y \qquad z = \tau \sinh \eta
\ee
one finally gets:
\bea\label{angmom4}
 J^y = && \tau_0\!\! \int\!\! \di x \, \di y \, \di \eta\; x \left[ 
- \cosh \eta \, (\varepsilon + p)  u^0 u^z \right. \nonumber \\ 
 && \left. + \sinh \eta \, [(\varepsilon +p) u^{z} u^z + p)] \right]
\eea
where: 
\bea\label{transf2}
  u^0 &=& \cosh \eta \; u^\tau + \tau \sinh \eta \; u^\eta \nonumber \\
  u^z &=& \sinh \eta \; u^\tau + \tau \cosh \eta \; u^\eta
\eea
being $u^\tau = \sqrt{1 + \tau^2 u^{\eta 2}}$. In the case of Bjorken initial conditions 
with $u^\eta=0$ and $u^\tau=1$, the eq.~(\ref{angmom4}) boils down to:
\be\label{angmom5}
 J^y = - \tau_0 \int \di x \, \di y \, \di \eta\; \varepsilon(x,y,\eta) \, x \,
 \sinh \eta
\ee
\\

\section*{APPENDIX B - Causality constraints}

The inequality expressing the causality constraint in the hydrodynamical picture 
of relativistic heavy ion collisions is that the initial longitudinal flow velocity must 
not exceed the velocity of beam protons $v_z < v_{\rm beam}$ (assuming vanishing 
initial transverse velocity):
\be\label{ineq1}
 |y| = \Big| \frac{1}{2} \log\frac{1+v_z}{1-v_z} \Big| = 
 | \log(u^0 + u^z) | \le y_{\rm beam}
\ee 
By using the transformation rules (\ref{transf2}):
\bea
 && \log(u^0 + u^z) = \log \left[ (\cosh\eta+\sinh\eta)(u^\tau+\tau u^\eta) \right] 
  \nonumber \\
 && = \log \left[ \e^\eta (\sqrt{1+\tau^2 u^{\eta2}} + \tau u^\eta) \right] \nonumber \\
 && = \eta + \log (\sqrt{1+\tau^2 u^{\eta2}} + \tau u^\eta) \nn
 && = \eta + {\rm asinh} (\tau u^\eta) \le y_{\rm beam}
\eea 
the inequality (\ref{ineq1}) becomes:
$$
 | \eta + {\rm asinh} (\tau u^\eta) | \le y_{\rm beam}
$$
which can be solved for $u^\eta$:
\be
  -\frac{1}{\tau} \sinh (y_{\rm beam} + \eta) \le 
  u^\eta \le \frac{1}{\tau} \sinh (y_{\rm beam}-\eta) 
\ee
The form (\ref{ueta}) fulfills the above inequality.

\end{document}